\documentclass[review,12pt]{elsarticle}

\usepackage{lineno}

\journal{Computers and Fluids}

\usepackage{graphicx}
\usepackage{amsfonts}
\usepackage{amssymb}
\usepackage{float}
\usepackage{latexsym}
\usepackage{amsmath}
\usepackage{subfigure}
\usepackage[euler]{textgreek}
\usepackage{textcomp}

\usepackage{algorithm}
\PassOptionsToPackage{noend}{algpseudocode}
\usepackage[end]{algpseudocode}

\usepackage[table]{xcolor}

\newcommand{\tcr}[1]{\textcolor{black}{#1}}

\newcommand{\eqr}[1]{(\ref{eq:#1})}

\newcommand{\pd}[2]{\frac{\partial #1}{\partial #2}}

\newcommand{\eg}{\textit{e.g.}}
\newcommand{\bs}[1]{\boldsymbol{#1}}
\newcommand{\etal}{\textit{et al.~}}
\newcommand{\ie}{\textit{i.e.}}

\graphicspath{ {./} }

\begin{document}

\begin{frontmatter}

\title{Direct numerical simulation of compressible interfacial multiphase flows using a mass-momentum-energy consistent volume-of-fluid method}

\author[add1]{Bo Zhang}
\author[add1]{Bradley Boyd}
\author[add1]{Yue Ling\corref{cor1}}
\ead{Stanley\_Ling@baylor.edu}
\cortext[cor1]{Corresponding author. 
	Address:  
 	One Bear Place \#97356, Waco, TX 76798
 }
\address[add1]{Department of Mechanical Engineering, Baylor University, Waco, Texas 76798, United States}

\begin{abstract}
Compressible interfacial multiphase flows (CIMF) are essential to different applications, such as liquid fuel injection in supersonic propulsion systems. Since high-level details in CIMF are often difficult to measure in experiments, numerical simulation is an important alternative to shed light on the unclear physics. A direct numerical simulation (DNS) of CIMF will need to rigorously resolve the shock waves, the interfaces, and the interaction between the two. A novel numerical method has been developed and implemented in the present study. The geometric volume-of-fluid (VOF) method is employed to resolve the sharp interfaces between the two phases. The advection of the density, momentum, and energy is carried out consistently with VOF advection. To suppress spurious oscillations near shocks, numerical diffusion is introduced based on the Kurganov-Tadmor method in the region away from the interface. The contribution of pressure is incorporated using the projection method and the pressure is obtained by solving the Poisson-Helmholtz  equation, which allows the present method to handle flows with all Mach numbers. The present method is tested by a sequence of CIMF problems. The simulation results are validated against theories, experiments, and other simulations, and excellent agreement has been achieved. In particular, the linear single-mode Richtmyer-Meshkov instabilities with finite Weber and Reynolds numbers are simulated. The simulation results agree very well with the linear stability theory, which affirms the capability of the present method in capturing the viscous and capillary effects on shock-interface interaction. 
\end{abstract}


\end{frontmatter}


\section{Introduction}
Compressible interfacial multiphase flows (CIMF) are encountered in a wide variety of applications, such as lithotripsy, raindrop damage in supersonic flight, and liquid fuel injection in supersonic propulsion systems. Direct numerical simulations (DNS) that can fully resolve the interfacial dynamics and instability, the shock-interface interaction, and the interfacial topology changes are essential to the investigation of CIMF, since they can shed light on the unclear flow physics that are hard to diagnose in experiments. Though DNS is feasible only for small-scale CIMF problems, the high-fidelity simulation data are important to the development of sub-scale physics-based or data-driven models, to enable accurate simulations of CIMF on larger scales. To fully resolve CIMF, the governing conservation laws must be solved by numerical methods that can well capture the sharp interface, the shock waves, and the interaction between them. It is essential to conserve mass of each phase. Furthermore, the surface tension on interfaces must be modeled and calculated rigorously. 

The challenges of resolving shock waves and contact discontinuities are rooted in the numerical oscillations generated due to the Gibbs phenomenon \citep{Leveque_2002a}. Therefore, a common feature of the various shock-capturing methods developed in the past is to suppress the spurious oscillations near the shocks without contaminating the regions with smooth flow properties. Originating from the pioneering work of Godunov \citep{Godunov_1959a}, many finite-volume shock-capturing methods are based on the exact or approximate solution of the Riemann problems \cite{Roe_1981a, Harten_1981a, Toro_1994a}. There is another family of shock-capturing methods that are Riemann-solver free \cite{Lax_1954a, Rusanov_1961a, Kurganov_2000a}, such as the central scheme of Kurganov and Tadmor \cite{Kurganov_2000a}, which can also produce high-resolution results near the discontinuities similar to the Riemann-solver methods. A comprehensive review of the shock-capturing methods can be found in the texts and reviews \citep{Leveque_2002a} and thus will not be repeated here. 

To capture the interfaces separating different fluids or phases, different interface-tracking methods have been developed. The methods can be in general separately as diffused-interface and sharp-interface methods. While the diffused-interface methods, such as the level-set method \citep{Osher_1988a, Sussman_1999a, Osher_2001a, Saurel_2018a, Jain_2020a} are often easier to implement, it is often hard to guarantee exact mass conservation for each phase in highly deforming flows. In contrast, the sharp-interface methods, such as the arbitrary Lagrangian-Eulerian method \citep{Hu_2001a, Luo_2004a,Corot_2020a}, the front-tracking method \citep{Unverdi_1992a, Tryggvason_2001a, Bo_2011a}, and the volume-of-fluid (VOF) method \citep{Hirt_1981a, Lafaurie_1994a, Scardovelli_1999a}, are more complicated, but they are usually more accurate, in particular on conserving mass. For CIMF that involve interfacial topology changes, such as atomization, the level-set and VOF methods are particularly popular due to their capabilities to handle topology changes. The VOF method also has the important feature in conserving mass. The conventional level-set methods suffer from not conserving mass, and usually need to be used together with the VOF method \cite{Sussman_2000a} . 

The dynamics of the interface is influenced by motion of the fluids on both sides and also the surface tension. Accurate calculation of surface tension is thus critical, no matter which interface-capturing method is to be used \citep{Popinet_2018a}. In some previous simulations of CIMF, the surface tension is ignored since the time scale of interest is much smaller than the capillary time scale. For such cases, it is acceptable to ignore surface tension  \cite{Johnsen_2006a, Meng_2015a, Meng_2018a}. However, for CIMF that involve small interfacial length scales or topology changes due to liquid breakups, the capillary time scale becomes comparable to the flow time scale, then the surface tension is important and must be rigorously incorporated in the simulation. The relative importance of viscosity compared to surface tension is characterized by the Ohnesorge (Oh) number, for problems with low Oh, the viscous effects can be ignored in the simulations \cite{Schmidmayer_2017a,Corot_2020a}. Different modeling approaches for surface tension can be found in the review by Popinet \cite{Popinet_2009a}. The continuous surface force (CSF) \cite{Brackbill_1992a} approach is typically used in CIMF  \cite{Chauveheid_2015a, Fuster_2018a}. A well-known issue of surface tension calculation based on the VOF method is the spurious or parasitic currents \cite{Renardy_2002a}, which arise in simulation of a droplet in equilibrium. The balanced-force discretization method was developed to resolve this issue, in which the discretization of surface tension is consistent with that for pressure \cite{Francois_2006a, Popinet_2009a}. The interface curvature is required for computing the surface tension. The evaluation of curvature is relatively easy for the level-set methods since the interface is a continuous function, but the task becomes more complex for the VOF methods since the volume fraction jumps across the interface. A possible solution is the convolution method which uses a smoothed volume fraction \cite{Brackbill_1992a}. Nevertheless, Afkhami and Bussman \cite{Afkhami_2008a} showed that the convolution method yields poor accuracy and the height-function method, which is based on the heights of the interface in a local coordinate, is a better solution. 

Numerical modeling and simulation of CIMF with surface tension are challenging. \tcr{Except for a few pioneering works \cite{Perigaud_2005a}, most studies in the literature emerged in the past decade \cite{Jemison_2014a, Rohde_2015a, Chauveheid_2015a, meng2016numerical, Schmidmayer_2017a, Garrick_2017a, Garrick_2017b, Fechter_2018a, Fuster_2018a, Arienti_2019a, Corot_2020a, Jain_2020a, Oomar_2021a}. Conventionally, the advection and pressure terms in the momentum and energy equations are coupled and solved with explicit time integration, and approximate Riemann solvers (\ie, HLLC) were used for the calculation of inviscid fluxes in the conservation laws and the advection equation for the indicator function or volume fraction for the reference phase \cite{Perigaud_2005a, meng2016numerical, Rohde_2015a}. There are a couple of numerical issues for such approaches. First of all, the numerical diffusion will lead to a diffused interface. To avoid excessive smearing at the interface, the interface compression technique is required \cite{Garrick_2017b}. Another option is to solve the advection equation for the volume fraction using algebraic or geometric VOF methods \cite{Xiao_2005a,Garrick_2017a, Jemison_2014a, Arienti_2019a, Oomar_2021a} or using the Ghost-Fluid method \cite{Fechter_2018a}.  While VOF or Ghost-Fluid methods maintain sharp-interface tracking, often there is an inconsistency between the advection of the interface and the conservative variables (\ie, momentum). This inconsistency may cause numerical instability when there is a large contrast in the material properties across the interface \cite{Arrufat_2020a, Zhang_2020a}}. The second issue is the acoustic time step restriction. In some applications of CIMF, the Mach number varies significantly in phases and in space, \eg, the Mach number is high in the gas but is low in the liquid. If the pressure is coupled with the advection of the conservative variables, then the time step must be smaller than the minimum acoustic time step in the domain. 

The all-Mach methods have been shown to be successful in alleviating the acoustic time step constraint for single-phase compressible flows, see \eg, \cite{Kwatra_2009a}, which were then extended to multiphase flows by Jemison \etal \cite{Jemison_2014a} and Fuster and Popinet (FP) \cite{Fuster_2018a}. In these methods, the advection and pressure terms are calculated separately. On one hand, the advection terms in the conservation laws are treated explicitly and the time step only needs to respect the fluid time scale. On the other hand, the pressure terms in the momentum and energy equations are treated implicitly, so that the acoustic time step restriction can be lifted. The additional Poisson equation is solved for pressure, which is then used to correct the momentum and energy. Furthermore, Jemison \etal \cite{Jemison_2014a} showed that such semi-implicit methods are asymptotic preserving, namely the incompressible pressure projection method is recovered in the asymptotic limit of infinite sound speed, and this feature is important for flows with a large contrast in material properties.

Another important feature for the methods by Jemison \etal \cite{Jemison_2014a}, \tcr{Arienti \etal \cite{Arienti_2019a}}, and Fuster and Popinet (FP) \cite{Fuster_2018a} is that the advection of the conservative variables, including mass, momentum, and energy for each phase is consistent with that of the volume fraction of the corresponding phase. For advection of the volume fraction, Jemison \etal \cite{Jemison_2014a} and \tcr{Arienti \etal \cite{Arienti_2019a}} used the moment-of-fluid (MOF) method, while Fuster and Popinet (FP) \cite{Fuster_2018a} used the VOF method. The difference is that MOF method uses the centroid of the reference phase to determine the orientation of the interface. In the method of Jemison \etal \cite{Jemison_2014a}, the advection flux for a conservative variable is computed by integrating it over the volume of the corresponding phase to be advected, based on the linear reconstructed interface and the field of the variable. The FP method has used a more convenient approach, where the advection of the conservative variables for each phase is achieved by advecting them as tracers associated with the volume fraction for the corresponding phase.

The present study aims at extending the FP method to enable direct numerical simulation of CIMF. Two extensions have been made. 
\tcr{First, additional numerical diffusion is introduced to eliminate the spurious oscillations near the shocks observed in the FP results.  Similar numerical oscillations have been observed in the results of Jemison \etal  \cite{Jemison_2014a}, which are more profound when the time step is small. The advection fluxes in the FP method are calculated as the product of the conservative variables to be advected and the cell surface velocity. While the former is computed based on the Bell-Corella-Glaz upwind scheme, the latter is approximated using the central differencing. The pressure terms (in momentum and energy equations) are discretized by central differencing as well. As a result, the numerical diffusion induced by the overall inviscid fluxes is not sufficient to damp the oscillations. The additional numerical diffusion is computed following the central upwind method of Kurganov and Tadmor (KT) \citep{Kurganov_2000a, Kurganov_2001a, Kurganov_2002a}. }
Second, the contribution of the viscous dissipation to the pressure evolution equation, which was ignored by FP, is incorporated.
Eventually, the present method exhibits important features, including 1) accurately capturing the sharp interface through the geometric VOF method; 2) consistent advection of conservative variables at the interface; 3) effective elimination of numerical oscillations induced by shock and discontinuities; 4) rigorous incorporation of surface tension and viscosity, which are essential to  accurate simulation of shock-interface interaction when viscous and surface tension effects are important.

Finally, a comprehensive test suite will be established to validate the present method in capturing shock-interface interaction and the resulting interfacial dynamics and instability. The remainder of the manuscript is organized as follows. The governing equations are presented in section \ref{sec:model}. The numerical methods are introduced in section \ref{sec:method}. The test results will be presented and discussed in section \ref{sec:results}. Finally, the conclusions will be drawn in section \ref{sec:conclusions}.

\section{Governing Equations} 
\label{sec:model}
\subsection{Conservation laws}
The gas and liquid phases in the compressible interfacial multiphase flows satisfy the conservation laws for mass, momentum and energy, 
\begin{align}
	\pd{\rho_k}{t} + \pd{\rho_k u_{k,i}}{x_i} & = 0 \, ,
 	 \label{eq:continuity1}\\
	\pd{\rho_k u_{k,i}}{t} + \pd{\rho_k u_{k,i}u_{k,j}}{x_j} & = -\pd{p_k}{x_i} + \pd{\tau_{k,ij}}{x_j} \, ,
	\label{eq:mom} \\
	\pd{E_k}{t} + \pd{E_k u_{k,i}}{x_i} & = -\pd{p_k u_{k,i}}{x_i} + \pd{\tau_{ij}u_{k,i}}{x_i} - \pd{q_{k,i}}{x_i}\, ,
	\label{eq:energy1}\,
\end{align}
where the subscript $k=l,g$ denotes the liquid ($l$) and the gas ($g$) phases, respectively. Furthermore, $\rho_k$, $u_k$, and $p_k$ represent density, velocity, and pressure. The total energy is denoted by $E_k=\rho_k(e_k + \frac{1}{2} u_{k,i}u_{k,i})$, where $e_k$ is the internal energy. The viscous stress tensor in each phase is represented as $\tau_{k,ij} = \mu_k(\partial_j{u_{k,i}}+ \partial_i{u_{k,j}}) +\lambda_{k,v}\partial_k{u_{k,k}}\delta_{ij}$, where $\mu_k$ and $\lambda_{k,v}$ are the coefficients of viscosity. The heat flux in each phase is represented by $q_{k,i}$, though in the present study the thermal diffusion is ignored. 

The internal energy and pressure are related by the equation of state (EOS). The stiffened EOS in the Mie-Gr\"uneisen form is commonly used, 
\begin{align}
  \rho_k e_k& = \frac{p_k + \gamma_k \Pi_{k,\infty}}{\gamma_k - 1} \, ,
\end{align}
where $\gamma_k$ is the specific heat ratio and $\Pi_{k,\infty}$ is the reference pressure for each phase. When $\Pi_{\infty,k}=0$, the Mie-Gr\"uneisen EOS reduces to the ideal gas EOS. 
The values of $\gamma_k$ and $\Pi_{\infty,k}$ for a given material are obtained by fitting the corresponding shock compression experimental data \citep{Marsh_1980a}. The speed of sound can be computed as 
\begin{equation}
	c_k = \sqrt{\frac{\gamma_k (p_k + \Pi_{\infty,k})}{\rho_k}}\,.
\end{equation}

The present numerical framework can also accommodate other EOS, such as the Jones-Wilkins-Lee (JWL) EOS: 
\begin{align}
  p_k = \Gamma_k \rho_k e_k + f(\rho)\, , 
\end{align}
where 
\begin{align}
 f(\rho) = A_k \left( 1- \frac{\Gamma_k}{R_{1,k}}\frac{\rho_{0,k}}{\rho_k}\right) 
  \exp\left(-R_{1,k}\frac{\rho_{0,k}}{\rho_k}  \right) \nonumber \\
  + B_k \left( 1- \frac{\Gamma_k}{R_{2,k}}\frac{\rho_{0,k}}{\rho_k}\right) 
  \exp\left(-R_{2,k}\frac{\rho_{0,k}}{\rho_k}  \right)
\end{align}
is the correction term for high-pressure gases. The model constants for phase $k$ include 
the reference density $\rho_{0,k}$, the low-pressure Gr\"uneisen coefficient $\Gamma_k$, 
the high-pressure coefficients $A_k$ and $R_{1,k}$, and the intermediate-pressure coefficients 
$B_k$ and $R_{2,k}$. 
The sound speed for the JWL EOS is expressed as 
\begin{equation}
	c_k^2= (\Gamma_k +1) p_k/\rho_k - f(\rho_k)/\rho_k - f'(\rho_k)/\rho_k^2\,.
\end{equation}

\subsection{Interfacial conditions}
The two different phases are distinguished by a characteristic function $\chi$. Generally we use $\chi=1$ and $0$ to represent the liquid and gas phases, respectively. The advection equation for $\chi$ is given as
\begin{align}
  \pd{\chi}{t} + u_i \pd{\chi }{x_i} & = 0 \, .
  \label{eq:adv}
\end{align}

The fluid properties jump across the interface separating the two phases. While the velocity is the same across the interface, there exists a jump in the stress due to the contribution of surface tension, 
\begin{align}
  \big[u_i\big]_s=0\,, \quad -\big[-p + n_i\tau_{ij}n_j\big]_s=\sigma \kappa\,,
\end{align}
where $[\cdot ]_s$ represents the jump of a variable across the interface.  The surface tension coefficient $\sigma$ is taken to be constant in the present study. The local curvature and the unit normal vector of the interface are denoted by $\kappa$ and $n_i$, respectively. 

\subsection{Model equations}
The mean value of $\chi$ in a computational cell is defined as 
\begin{align}
	f = \frac{1}{\Delta\Omega}\int_{\Omega} \chi dV\, .
	\label{eq:liq_vof}
\end{align}
which also represents the volume fraction of liquid ($\chi=1$) in a cell. Similarly, $\hat{f}=1-f$ is the gas volume fraction.  

While the conservation laws for each phase are satisfied in the cells fully occupied by liquid or  gas, \ie, $f=1$ or 0, additional modeling efforts are required for the interfacial cells ($0<f<1$), which contain a liquid-gas ``mixture". 

The volume-average properties for the gas-liquid mixture, denoted by variables without a subscript. The volume-average density, pressure, and total energy of the mixture are expressed as
\begin{align}
	\rho = f \rho_l + \hat{f} \rho_g\,, \quad
	p =fp_l + \hat{f}p_g\,, \quad
	E = f E_l + \hat{f}E_g\, . 
	\label{eq:mix_density}
\end{align}
Similarly the mixture momentum is defined as
\begin{align}
	\rho u_i & = f \rho_l u_i + \hat{f} \rho_g u_i\,.
	\label{eq:mix_kom}
\end{align}
Since the velocity is continuous at the interface, it is unnecessary to distinguish the gas and liquid velocities. The pressure on the liquid and gas sides are different due to the surface tension. In cells with $f=1$ or $f=0$, $p$ is identical to the liquid pressure $p_l$ or the gas pressure $p_g$. In interfacial cells, $p$ represent the average pressure, then the pressure on the liquid and gas sides of the interface can be calculated by the Laplace relation, namely 
\begin{align}
	p_l & =  p + \hat{f} \sigma \kappa
  \label{eq:pl}\\
	p_g & =  p - f \sigma \kappa \,.
 \label{eq:pg}
\end{align} 

Incorporating the above mixture rules, the governing equations for the two-phase model can be written as
\begin{align}
	\pd{f \rho_l}{t} + \pd{f\rho_l u_i}{x_i} & = 0 \, ,
	\label{eq:mass_liq}\\ 
	\pd{\hat{f} \rho_g}{t} + \pd{\hat{f} \rho_g u_i}{x_i} & = 0 \, ,
	\label{eq:mass_gas} \\
	\pd{f \rho_l u_i}{t} + \pd{f \rho_l u_iu_j}{x_j} & = - f \pd{p}{x_i} + f \pd{\tau_{ij}}{x_j} - f \sigma \kappa \pd{\hat{f}}{x_i}\, ,	
	\label{eq:mom_liq}\\
	\pd{\hat{f} \rho_g u_i}{t} + \pd{\hat{f} \rho_g u_iu_j}{x_j} & = - \hat{f} \pd{p}{x_i} + \hat{f} \pd{\tau_{ij}}{x_j} + \hat{f} \sigma \kappa \pd{{f}}{x_i}\, ,	
	\label{eq:mom_gas}\\
	\pd{fE_l}{t} + \pd{fE_l u_i}{x_i} & = -f\pd{p u_i}{x_i} + f \pd{\tau_{ij}u_{i}}{x_j} + f \sigma \kappa \pd{\hat{f} u_i}{x_i}\, ,
	\label{eq:energy_liq}\\
	\pd{\hat{f}E_g}{t} + \pd{\hat{f}E_g u_i}{x_i} & = -\hat{f}\pd{p u_i}{x_i} + \hat{f}\pd{ \tau_{ij}u_{i}}{x_j}- \hat{f} \sigma \kappa \pd{f u_i}{x_i}\, .
	\label{eq:energy_gas}
\end{align}

The momentum and energy equations for the mixture can be obtained by summing Eqs.\ \eqr{mom_liq}-\eqr{mom_gas}, and Eqs.\ \eqr{energy_liq}-\eqr{energy_gas}, respectively, 
\begin{align}
	\pd{\rho u_i}{t} + \pd{\rho u_iu_j}{x_j} & = -\pd{p}{x_i} + \pd{\tau_{ij}}{x_j} + \sigma \kappa \pd{f}{x_i}\, ,	
	\label{eq:mom_2}\\
	\pd{ E}{t} + \pd{ E u_i}{x_i} & = -\pd{p u_i}{x_i} + \pd{\tau_{ij} u_i }{x_j} + \sigma \kappa u_i \pd{f}{x_i}\, .	
	\label{eq:energy_2}
\end{align}
The internal energy equation for the mixture can be obtained by subtracting the kinetic energy portion from Eq.\ \eqr{energy_2}
\begin{align}
 	\rho \bigg(\pd{e}{t} + u_i \pd{e}{x_i}\bigg) =  -p \pd{u_i}{x_i} + \Phi_{v} \, ,
 \label{eq:internal_energy} 
\end{align}
where $\Phi_v=\tau_{ij} \pd{u_i}{u_j}$ is the viscous dissipation and 
the mixture internal energy is defined as $e = {E}/{\rho} - {u_i u_i}/{2}$. 
 Note that the surface tension has no contribution to the internal energy. The internal energy equation can be rewritten in terms of pressure as
\begin{align}
 \frac{1}{\rho c_\text{eff}^2} \bigg(\pd{p}{t} + u_i \pd{p}{x_i}\bigg) - \frac{\beta_T \Phi_{\nu}}{\rho C_p} & = -\pd{u_i}{x_i} \, ,
 \label{eq:press_evol} 
\end{align}
where $\beta_T$ and $C_p$ are the the thermal expansion coefficient and specific heat for constant pressure. 
The effective sound speed $c_\text{eff}$ is defined as 
\begin{align}
 \frac{1}{\rho c_\text{eff}^2}= \frac{\gamma}{\rho c^2} - \frac{\beta_T^2 T}{\rho C_p}\, ,
\end{align}
where $T$ is the temperature and $C_p$ is the specific heat for constant pressure. It can be approximated that $1/\rho c_\text{eff}^2 \approx 1/\rho c^2$ \cite{Fuster_2018a}.

\section{Numerical Methods}
\label{sec:method}
The governing equations listed in section \ref{sec:model} are solved by the finite volume approach on a collocated grid. The advection equation for the characteristic function, Eq.\ \eqr{adv} is discretized and solved using the VOF method. Following the FP method \cite{Fuster_2018a}, the advection terms for mass, momentum, and energy equations are handled for each phase separately, based on Eqs.\ \eqr{mass_liq}-\eqr{energy_gas}. To reduce the spurious oscillations near the interfaces, artificial numerical diffusion is introduced following the the central method of Kurganov and Tadmor \cite{Kurganov_2000a}. The viscous and surface tension terms for the momentum are incorporated using the mixture momentum equation Eq.\ \eqr{mom_2}, while those for the energy are computed for each phase separately by Eqs.\ \eqr{energy_liq} and \eqr{energy_gas}. To allow an all-Mach capability, the Helmholtz-Poisson equation is solved for the pressure, which is derived from the mixture internal-energy equation in terms of mean pressure (Eq.\ \eqr{press_evol}). The obtained pressure will be used to correct the velocity and the energy for each phase.  The variables that are eventually solved in each time step include $f\rho_l, \hat{f}\rho_g, fE_l, \hat{f}E_g,\rho u_i$, and $p$. The detailed procedures are described below. 

\subsection{VOF Advection}
The advection equation (Eq.\ \eqr{adv}) is first cast into its conservative form
\begin{align}
\pd{\chi}{t}+ \pd{(\chi u_i)}{x_i} = \chi \pd{u_i}{x_i} \, ,
\label{eq:char_func}
\end{align}
which can be integrated in a computational cell as 
\begin{align}
\Delta \Omega \pd{f}{t}+\oint_{\partial \Omega} \chi u_i n_i \mathrm{d}s =\int_{\Omega} \chi\pd{u_i}{x_i}\mathrm{d} V\, ,
\label{eq:adv_color_func1}
\end{align}
where $\Delta\Omega$ and $\partial \Omega$ represent the volume and the surface of the cell, and $f$ is the cell average of $\chi$, as defined in Eq.\ \eqr{liq_vof}. 

The volume fraction advection equation is then discretized in a direction-split form as follows,
\begin{align}
\Delta \Omega \frac{f^{n+1}- f^{n}}{\Delta t}+\Delta_i F_{f,i} = \chi_c \pd{u_i}{x_i} \Delta \Omega\, ,
\label{eq:adv_color_func}
\end{align}
where $\chi_c$ is the value of $\chi$ at the cell center (evaluated as $\chi_c = 1$ if $f > 0.5$ and $\chi_c = 0$ if $f \le 0.5$). As demonstrated by Weymouth and Yue \citep{Weymouth_2010a},  $\chi_c$ must be kept constant for all sweeping directions to ensure the exact mass conservation. The sum of net fluxes in all directions is denoted by $\Delta_i F^n_{f,i}=\Delta_x F^n_{f,x}+\Delta_y F^n_{f,y}+\Delta_z F^n_{f,z}$. The net flux in the $x$ direction is computed as 
\begin{align}
	\Delta_x F^n_{f,x} = F^n_{f,j+1/2} - F^n_{f,j-1/2}
\end{align}
where the flux of $F^n_{f,j+1/2}$ on a cell surface is computed as 
\begin{align}
F^n_{f,j+1/2}= f^n_{a,j} u^n_{f, j+1/2}S_{j+1/2} \,,
\label{eq:VOF_flux}
\end{align}
where $f^n_{a,j}$ represents the volume fraction of liquid in the cell $i$ to be advected across the cell surface within $\Delta t$, and $S_{j+1/2}$ is the cell surface area. The interface is first reconstructed by the piecewise linear interface construction (PLIC) method, where the interface normal is computed by the Mixed-Youngs-Centered (MYC) method \cite{Aulisa_2007a}. Then $f^n_{a,j}$ is evaluated based on the geometric reconstruction \citep{Scardovelli_1999a}. The $x$-velocity at the cell surface $u^n_{f, j+1/2}$ is approximated by central differencing from the values of the neighboring cells.

\subsection{Consistent and conservative advection of conservative variables}
It has been demonstrated in previous studies that solving the mass (VOF) and momentum equations consistently is critical to yield accurate results for two-phase flow with large density contrast \cite{Rudman_1998a, Chenadec_2013a, Vaudor_2017a, Zhang_2020a}. Therefore, the discretization of advection terms in the governing equations (Eqs.\ \eqr{mass_liq}-\eqr{energy_gas}) is consistent with the VOF method (Eq.\ \eqr{adv}). The conservative variables for each phase are advected as tracers associated with the volume fraction of the corresponding phase non-diffusively \citep{Lopez-Herrera_2015a}. Similar to the VOF method, the discretization for the convection terms is  conservative. 

Since the momentum is stored and solved in the mixture form, the liquid and gas momentum need to be calculated before they are advected, 
\begin{align}
	[f(\rho u_i)_l]^n & = (f \rho_l)^n u_i^n\,,\\ 
	[\hat{f}(\rho u_i)_g]^n & = (\hat{f} \rho_g)^n u_i^n\,,
\end{align}
where 
\begin{align}
	u_i^n & = (\rho u_i)^n /[f \rho_l)^n + (\hat{f} \rho_g)^n]\,.
	\label{eq:mix_velocity}
\end{align}

The fluxes for the conservative variables for the liquid, \ie, $\bs{U}^n_l=[f\rho_l, f(\rho u_i)_l, fE_l]^n$ and for the gas, \ie, $\bs{U}^n_l=[\hat{f}\rho_g, \hat{f}(\rho u_i)_g, \hat{f}E_g]^n$, are then computed based on the $f$ and $\hat{f}$ fluxes, 
\begin{align}
	\bs{F}^n_{U_l,j+1/2} & = \bs{V}^n_{la,j} F^n_{f,j+1/2}\,,
	\label{eq:tracer_flux_liq}\\
	\bs{F}^n_{U_g,j+1/2} & = \bs{V}^n_{ga,j} F^n_{\hat{f},j+1/2}\,,
	\label{eq:tracer_flux_gas}
\end{align}
	where $F^n_{f,j+1/2}$ is the VOF flux for the liquid volume fraction $f$, given in Eq.\ \eqr{VOF_flux}, and the flux for the gas volume fraction $\hat{f}$ is given as 
\begin{equation}
	F^n_{\hat{f},j+1/2}=\hat{f}^n_{a,j} u^n_{f, j+1/2}S_{j+1/2}\, .
\end{equation}
The values of the conservative variables for a given phase to be advected across the cell surface, \ie, $\bs{V}^n_{la,j}=[\rho_l, \rho_l u_i, E_l]_{a,j}^n$ and $\bs{V}^n_{ga,j}=[\rho_g, \rho_g u_i, E_g]_{a,j}^n$, are computed by linear reconstruction of the corresponding variable within the $j$ cell based on the Bell-Colella-Glaz scheme \citep{Bell_1989a} and the minmod slope limiter. The detailed expressions can be found in  Ref.~\cite{Fuster_2018a}. 	

The conservative variables for each phase are then integrated over time similar to Eq.\ \eqr{adv_color_func}
\begin{align}
  \Delta \Omega \frac{\bs{U}_{l}^{*} - \bs{U}_{l}^n}{\Delta t} = \Delta_i \bs{F}^n_{U_{l},i}\,,\\ 
    \Delta \Omega \frac{\bs{U}_{g}^{*} - \bs{U}_{g}^n}{\Delta t} = \Delta_i \bs{F}^n_{U_{g},i}\,,
\label{eq:euler_advect}
\end{align}
where $\Delta_i \bs{F}^n_{U_l,i}$ and $\Delta_i \bs{F}^n_{U_g,i}$ are the sums of net fluxes in all directions and the superscript $*$ represent the updated variables after the convection step. The net fluxes in the $x$ direction are calculated as 
\begin{align}
	\Delta_x \bs{F}^n_{U_l,x} = \bs{F}^n_{U_l,j+1/2} - \bs{F}^n_{U_l,j-1/2}\,, \\
	\Delta_x \bs{F}^n_{U_g,x} = \bs{F}^n_{U_g,j+1/2} - \bs{F}^n_{U_g,j-1/2}\,. 
\end{align}

After computing the momentum for each phase, \ie, $f(\rho u_i)_l^*$ and $\hat{f}(\rho u_i)_g^*$, the mixture momentum $(\rho u_i)^*$ is updated using Eq.\ \eqr{mix_kom}.

\subsection{Numerical diffusion}
\label{sec:kt}
The  advection method described above for the conservative variables is essential to obtaining accurate results near the interface \cite{Fuster_2018a}. Nevertheless, as will be shown later in section \ref{sec:results}, the method induces numerical oscillations near discontinuities like shock waves and tailing edge of expansion fan. 
\tcr{As shown in Eqs.~\eqr{tracer_flux_liq} and \eqr{tracer_flux_gas}, the advection fluxes are calculated as the product of the conservative variables to be advected, $\bs{V}^n_{la,j}$ or $\bs{V}^n_{ga,j}$, and the VOF flux based on the cell surface velocity $u_f$. While $\bs{V}^n_{la,j}$ and $\bs{V}^n_{ga,j}$ computed based on the Bell-Corella-Glaz upwind scheme, $u_f$ is approximated using the central differencing. The pressure terms (in momentum and energy equations) are also discretized by central differencing. The numerical diffusion induced by overall inviscid fluxes (for advection and pressure terms) is not sufficient to damp the numerical oscillations. }
In order to eliminate these numerical oscillations, additional numerical diffusion is introduced based on the central-upwind method of Kurganov \etal \cite{Kurganov_2001a}. The overall flux in the central-upwind method can be decomposed to central-difference part and the numerical-diffusion part. Here, only the numerical-diffusion portion is employed, which is expressed as

\begin{align}
\bs{H}_{j+\frac{1}{2}} &= a \big[\bs{U}_{j+\frac{1}{2}}^{+} - \bs{U}_{j+\frac{1}{2}}^{-}\big] S\, , 
\label{eq:kurganov_flux}
\end{align}
where 
\begin{equation}
a = \frac{a_{j+\frac{1}{2}}^{+} a_{j+\frac{1}{2}}^{-}}{a_{j+\frac{1}{2}}^{+} - a_{j+\frac{1}{2}}^{-}}\, 
\label{eq:kurganov_speed0}
\end{equation}

and the superscripts $^+$ and $^-$ denote the fluid properties on the right and left sides of the cell surfaces, which are in turn obtained from linear reconstruction in the two neighboring cells as described in the advection step. The numerical diffusion is applied only in the cells without interfaces ($f=0$ or $f=1$), so there is no smearing of properties at the sharp interface. For convenience, we use the mixture notations $\bs{U}=[\rho, \rho u_i, E]$. Yet since the numerical diffusion is applied only to pure liquid and gas cells, $\bs{U}$ actually represent the $[f\rho_l, \rho u_i, fE_l]$ and $[\hat{f}\rho_g, \rho u_i, \hat{f}E_g]$ in cells with $f=1$ and $f=0$, respectively.  
The one-sided characteristic speeds are denoted as $a_{j+1/2}^{+}$ and $a_{j+1/2}^{-}$, which are calculated as \cite{Kurganov_2001a}
\begin{align}
a_{j+\frac{1}{2}}^{+} &= \text{max}\bigg(\lambda_{\max}\bigg(\pd{\bs{F}_{U,i}}{\bs{U}}(\bs{U}_{j+\frac{1}{2}}^{-})\bigg), \lambda_{\max}\bigg(\pd{\bs{F}_{U,i}}{\bs{U}}(\bs{U}_{j+\frac{1}{2}}^{+})\bigg), 0 \bigg)\,\label{eq:kurganov_speed1}
\\
a_{j+\frac{1}{2}}^{-} &= \text{min}\bigg(\lambda_{\min}\bigg(\pd{\bs{F}_{U,i}}{\bs{U}}(\bs{U}_{j+\frac{1}{2}}^{-})\bigg), \lambda_{\min}\bigg(\pd{\bs{F}_{U,i}}{\bs{U}}(\bs{U}_{j+\frac{1}{2}}^{+})\bigg), 0 \bigg) \, ,
\label{eq:kurganov_speed}
\end{align}
where $\lambda_{\min}$ and $\lambda_{\max}$ are the maximum and minimum eigenvalues of the Jacobian matrix $\partial{\bs{F}_{U,i}}/\partial{\bs{U}}$. When $a_{j+1/2}^{+}=-a_{j+1/2}^{-}=a_{j+1/2}$, $a=-a_{j+1/2}$ and Eq.\ \eqr{kurganov_flux} reduces to the form of for the classic KT method \cite{Kurganov_2000a}.

The discrete equation to update the conservative variables is
\begin{align}
   \Delta \Omega \frac{\bs{U}^{**} - \bs{U}^{*}}{\Delta t} &= \Delta_i \bs{H}^n_i\quad \text{if}\ f=0\ \text{or}\ f=1 \,, 
\end{align}
where $\Delta_i \bs{H}^n_i$ is the sum of net numerical-diffusion fluxes in all directions, and $\bs{U}^{**}$ represents the variables after the numerical-diffusion step. 

\subsection{Surface Tension and Viscous terms}
The viscous term in the momentum equation is discretized in time using the Crank-Nicholson method, while the surface tension term is treated explicitly. 
\begin{align}
	\frac{(\rho u_i)^{***} - (\rho u_i)^{**}}{\Delta t} -\frac{1}{2}  \pd{\tau^{***}_{ij}}{x_i}& =\frac{1}{2} \pd{\tau^n_{ij}}{x_i} +   \sigma \kappa\pd{( f)^n}{x_i}\, ,
	\label{eq:ST_kom} 
\end{align}
where the superscript $^{***}$ indicate the variables after incorporating the viscosity-surface-tension step. 

The central difference method is used to spatially discretize the viscous stress terms $\partial{\tau_{ij}}/\partial{x_i}$ and $\partial{\tau_{ij}u_i}/\partial{x_i}$. The viscosity at interfacial cells is computed by the arithmetic mean similar to density as $\mu=f\mu_l + \hat{f}\mu_g$. 

The balanced-force discretization method is used for the surface tension term, and the curvature $\kappa$ is computed by the height-function method \cite{Popinet_2009a, Francois_2006a}. Equation \eqr{ST_kom} is solved using a multigrid solver \cite{Popinet_2009a}.

The contributions for both the viscous stress and surface tension to the energy are incorporated as 
\begin{align}
	\frac{(fE_l)^{***} - (fE)_l^{**}}{\Delta t} & = f\pd{ \tau^n_{ij}u_i}{x_i} -  f \sigma \kappa\pd{(\hat{f} u_i)^n}{x_i}\,, 
	\label{eq:ST_energy_liq}\\
	\frac{(\hat{f}E_g)^{***} - (\hat{f}E)_g^{**}}{\Delta t} & = \hat{f}\pd{ \tau^n_{ij}u_i}{x_i} +  \hat{f}\sigma \kappa\pd{({f} u_i)^n}{x_i}\,. 
	\label{eq:ST_energy_gas}
\end{align}
Note that here we only include contribution of the Laplace pressure to the energy (see Eqs.\ \eqr{pl} and \eqr{pg}), the contribution for the mean pressure will be added after the projection step. 


\subsection{Poisson-Helmholtz Equation for Pressure}
\label{sec:project}
The contribution of pressure to the momentum and energy equation is incorporated using the projection method. 
The internal energy equation in terms of the mean pressure $p$, Eq.\ \eqr{press_evol}, can be discretized in time as
\begin{align}
	\frac{p^{n+1} - p^n}{\Delta t} + u_i^n\pd{p^n}{x_i} - \left(\frac{\beta_T c^2 \Phi_{\nu}}{ C_p}\right)^n = -(\rho c^2) \pd{u_i^{n+1}}{x_i} \, ,
  	\label{eq:discrete_press_evol}
\end{align}
where $u_i^{n+1}$ is the final velocity at $t^{n+1}$, which can be computed adding the contribution of pressure to $u_i^{***}$, 
\begin{align}
  u_i^{n+1} = u_i^{***} - \Delta t \left(\frac{1}{\rho} \pd{p}{x_i}\right)^{n+1} \, . 
  \label{eq:projection_vel}
\end{align}
Substitute Eq.\ \eqr{projection_vel} into \eqr{discrete_press_evol} and split the time integration of pressure in two steps, it yields
\begin{align}
  \frac{p^{n+1} - p^{***}}{\Delta t} &=-  (\rho c^2)^{***} \left(\pd{(u_i)^{***}}{x_i} + \Delta t \frac{\partial}{\partial{x_i}}\bigg(\frac{1}{\rho}\pd{p}{x_i}\bigg)^{n+1} \right)\, , 
  \label{eq:energy_discrete}    \\
  \frac{p^{***} - p^{n}}{\Delta t} & + u_i^n\pd{p^n}{x_i} - \left(\frac{\beta_T c^2 \Phi_{\nu}}{ C_p}\right)^n = 0\, ,
  \label{eq:prov_pressure}
\end{align}
where $p^{***}$ is a provisional pressure that accounts for only the convection and viscous terms. Therefore, instead of using Eq.\ \eqr{prov_pressure}, $p^{***}$ can be computed based on the conservative variables updated after the convection, numerical-diffusion, and the viscosity-surface-tension steps,  namely $[f\rho_l, \hat{f}\rho_g, \rho u_i, fE_l, \hat{f}E_g]^{***}$, using the EOS.

Since the interface separating the two immiscible fluids is considered as a sharp surface, there is no numerical mixing between the fluids. A ``mixture" rule is only required to calculate $p^{***}$ at the interfacial cells, where the volume fraction of the reference phase is fractional. When the stiffened EOS is used, the pressure can be calculated as
\begin{align}
  p^{***} = \left[{ \left((fE_l+ \hat{f}E_g)^{***}- \frac{1}{2} (\rho u_i u_i)^{***}\right)- {\frac{\gamma \Pi_{\infty}}{\gamma-1}}}\right]\left({{\frac{1}{\gamma-1}}}\right)^{-1}\, .
  \label{eq:prov_pressure_EOS}
\end{align}where 
\begin{align}
	{\frac{1}{\gamma-1}} &= \frac{f}{\gamma_l - 1} + \frac{\hat{f}}{\gamma_g - 1}\, ,
	\label{eq:mix_gamma}\\
	{\frac{\gamma \Pi_{\infty}}{\gamma-1}}& = \frac{f\Pi_{\infty,l} \gamma_l}{\gamma_l - 1} + \frac{\hat{f}\Pi_{\infty,g}\gamma_g}{\gamma_g - 1}.
	\label{eq:mix_Pi}
\end{align}

Equation \eqr{energy_discrete} can be recognized as the Poisson-Helmholtz equation, which is solved by the multigrid solver to obtain $p^{n+1}$. Introducing the Poisson-Helmholtz equation of pressure provides an important advantage of alleviating the acoustic time step restriction \cite{Kwatra_2009a}. 

Finally, the velocity is corrected by the new pressure using Eq.\ \eqr{projection_vel}, and similarly, the total energy for the gas and liquid are corrected as 
\begin{align}
	(f E_l)^{n+1} =(fE_l)^{***}- {\Delta t} f \pd{(p u_i)^{n+1}}{x_i} \, ,\\
	(\hat{f}E_g)^{n+1} =(\hat{f}E_g)^{***}- {\Delta t} \hat{f} \pd{(p u_i)^{n+1}}{x_i} \, .
  \label{eq:projection_energy}
\end{align}

\subsection{Time step and modified numerical diffusion}
It can be shown that the numerical diffusivity induced by the numerical diffusion flux defined in Eq.\ \eqr{kurganov_flux} is $\nu_{n}=-a\Delta x$. As the numerical diffusion is integrated in time explicitly, it will impose a time step constraint for numerical stability. The diffusion number for the numerical diffusion can be defined as 
\begin{equation}
	D_n = \frac{\nu_n \Delta t}{(\Delta x)^2} = \frac{-a \Delta t}{\Delta x}
	\label{eq:Diff_number}
\end{equation}
and $D_n<1/2$, 1/4, and 1/6 for diffusion equations in 1D, 2D, and 3D, respectively. For typical methods for compressible flows, the time step is constrained by the CFL condition based on the maximum velocity in the flows 
\begin{equation}
	C = \frac{V_{\max} \Delta t}{\Delta x} < 1/2\,.
\end{equation}
where $C$ is the Courant number. 
For high-Mach flows, $V_{\max}$ is the fluid velocity. For low-Mach flows $V_{\max}$ is dictated by the sound speed and the CFL condition becomes 
\begin{equation}
	C_a = \frac{c \Delta t}{\Delta x} < 1/2\, , 
\end{equation}
where $C_a$ is the acoustic Courant number.

If the time step is calculated based on $C_a$, then Eq.\ \eqr{Diff_number} becomes $D_n = {-a C_a}/{c}$. In the limit of zero Mach number, $a=-c$ and $D_n=C_a$, so if a small $C_a$ is used, then stability conditions for both the advection and numerical diffusion can be satisfied. However, for the present all-Mach approach, the time step can be larger than the acoustic time step. If a larger time step is used, \ie, $C_a>1/2$, then the stability condition for the numerical diffusion may not be satisfied. To guarantee stability, we propose to cap the numerical diffusivity by modifying Eq.\ \eqr{kurganov_speed0} as 
\begin{equation}
	a = - \min\left[ - \frac{a_{j+\frac{1}{2}}^{+} a_{j+\frac{1}{2}}^{-}}{a_{j+\frac{1}{2}}^{+} - a_{j+\frac{1}{2}}^{-}}, D_n\frac{\Delta x}{\Delta t} \right]\,.  
	\label{eq:kurganov_speed2}
\end{equation}
For a large $\Delta t$, the numerical diffusivity becomes $\nu_n=D_n (\Delta x)^2/\Delta t$, which will decrease as $\Delta t$ increases. In the results shown below, we use $D_n=0.1$ and calculate $\Delta t$ based on $C_a$. Another advantage of Eq.\ \eqr{kurganov_speed2} is that one can control the contribution of the numerical diffusion by varying $D_n$. For low-Mach flows without shock waves, the numerical diffusion is unnecessary and can be deactivated by setting $D_n$ to zero.

In the simulation, it is important to preserve positivity for physical properties like density. As demonstrated by Patkar \etal \cite{Patkar_2016a}, additional time step restrictions may be required to guarantee the positivity of the density and energy. These additional restrictions have not been used in the present study, nevertheless, for the tests performed, we have confirmed that the physical properties, including density, energy, pressure, and sound speed, are always positive.


\section{Test Cases and Results}
\label{sec:results}
The numerical methods described in section \ref{sec:method} are implemented in the open-source multiphase solver \emph{Basilisk} \cite{Basilisk}. A sequence of tests have been performed to validate the present method in resolving CIMF involving shock-interface interaction. The test cases are summarized in Table \ref{tab:cases}, along with the test purpose and validation approaches. Though the focus of the present study is on CIMF with surface tension, we have first performed tests without surface tension (sections \ref{sec:sp_shocktube}-\ref{sec:shock-bubble}), to examine the present methods in capturing the sharp interfaces without introducing spurious oscillations. The shock-droplet interaction problem in section \ref{sec:shock-drop} is employed to validate the present method in resolving multiphase flows with large contrast of properties across the interface. The Richtmyer-Meshkov instability with finite Reynolds and Weber numbers in section \ref{sec:rmi} is simulated to demonstrate the capability of the present methods in capturing shock-interface interaction with viscosity and surface tension. Finally, the capillary oscillation of a 2D drop is performed to show that the present method is valid in resolving low-Mach surface-tension driven flows with time steps larger than the acoustic time step.

\begin{table*}[tbp]
\centering
\begin{tabular}{c c c }
    \hline
Test 	& Test purpose	& Validation \\
	\hline
Single-phase shocktube	& Shock capturing	& Theory\\
Two-phase shocktube	& Shock \& interface capturing	& Theory\\
Shock-bubble	& Shock-interface interaction & Exp./Sim.\cite{Quirk_1996a}\\
Shock-droplet  & Shock-interface interaction & (2D) Exp.\cite{Igra_2001a}\\
 &  & (3D) Sim.\cite{Meng_2018a}\\
Linear RMI & Viscosity \& surface tension & Theory \cite{Carles_2002a}\\
Drop oscillation & Surface tension \& low-Mach flows & Theory/Sim.\cite{Perigaud_2005a}\\
    \hline
\end{tabular}
\caption{Summary of test cases.}
\label{tab:cases}
\end{table*}

\subsection{Single-phase Sod's shocktube}
\label{sec:sp_shocktube}
The 1D shocktube problem of Sod \citep{Sod_1978a} is a classic benchmark test case for shock-capturing schemes. Viscosity and surface tension are neglected in this test. The gas is taken to be ideal gas, so $\Pi_\infty=0$.

\begin{figure}[tbp]
\begin{center}
\includegraphics [width=1\columnwidth]{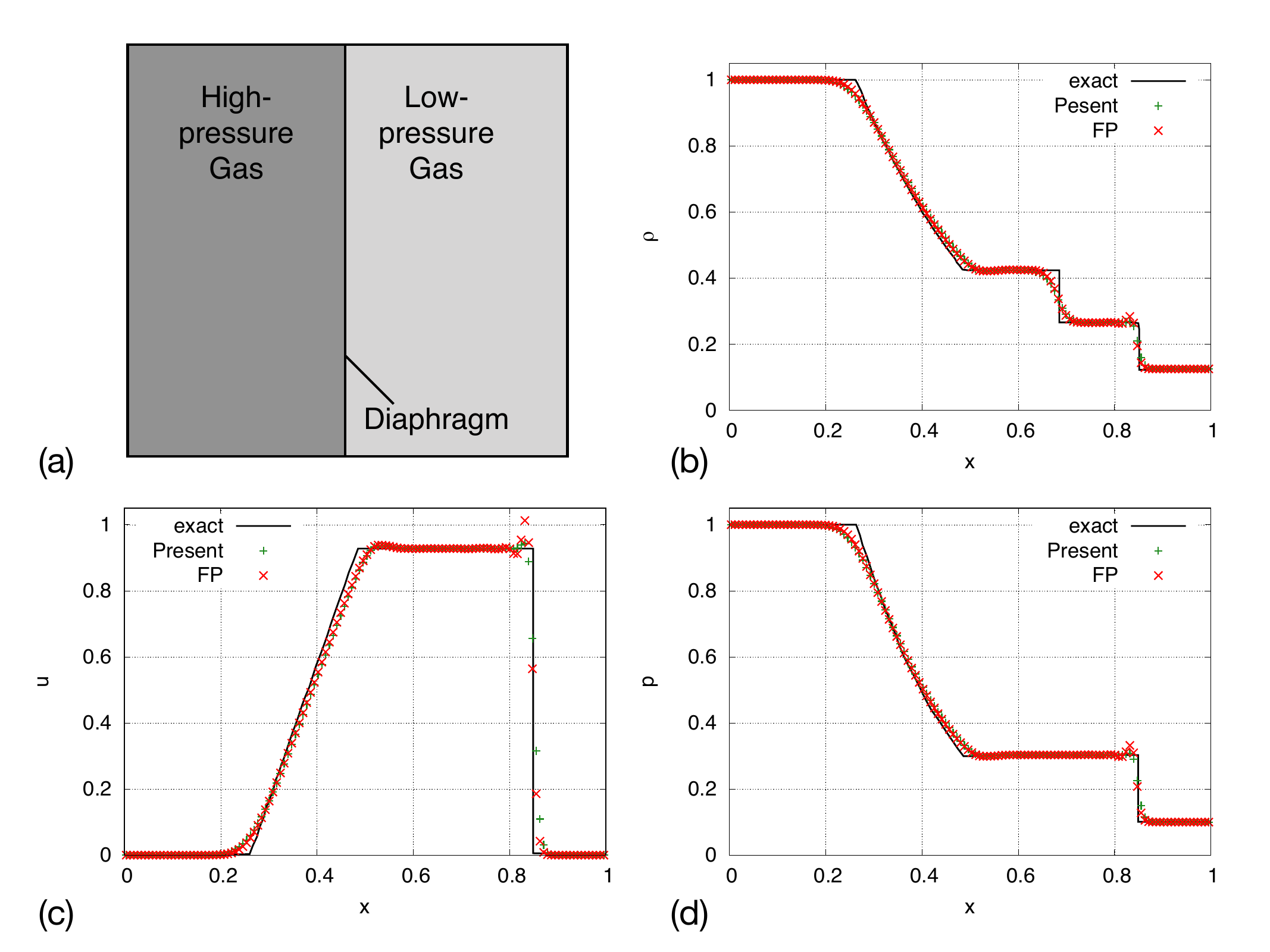}
\end{center}
\caption{Simulation setup (a) and results of (b) density, (c) velocity, and (d) pressure for the 1D shocktube problem at $t=0.2$. The solid lines represent the exact solution, and ``FP" represents the results using the method by Fuster and Popinet \cite{Fuster_2018a}. The cell size is $\Delta x=1/128$ and the time step is based on the acoustic Courant number $C_a=0.4$. }
\label{fig:shocktube_Sod}
\end{figure}

The same gas with high and low pressure is initially separated with a diaphragm, see Fig.\ \ref{fig:shocktube_Sod}(a). The initial conditions for the left and right states are given as
\begin{align}
 \{\rho, u, p, \gamma\} &=
 \begin{cases}
\{1.0, 0, 1.0, 1.4\} & 0 \le x \le 0.5, \\
  \{0.125, 0, 0.1, 1.4\} & 0.5 \le x \le 1.0.
 \end{cases}
 \label{eq:Sod_ST}
\end{align}
The domain is a square with the edge length $L=1$, which is discretized by a uniform mesh with a cell size $\Delta x=L/128$. At $t=0$, the diaphragm is located at $x=0.5$. As the diaphragm is removed,  the shock wave, the contact surface, and the expansion fan are generated. The numerical results at $t=0.2$ for the present and the FP methods for density, pressure and velocity are compared with the theory in Figs.\ \ref{fig:shocktube_Sod}(b)-(d). Results for both methods agree with the exact solution well in general. Spurious oscillations are observed near the shocks for the FP results, while the oscillations are effectively suppressed by the present method.

\begin{figure}[tbp]
\begin{center}
\includegraphics [width=1\columnwidth]{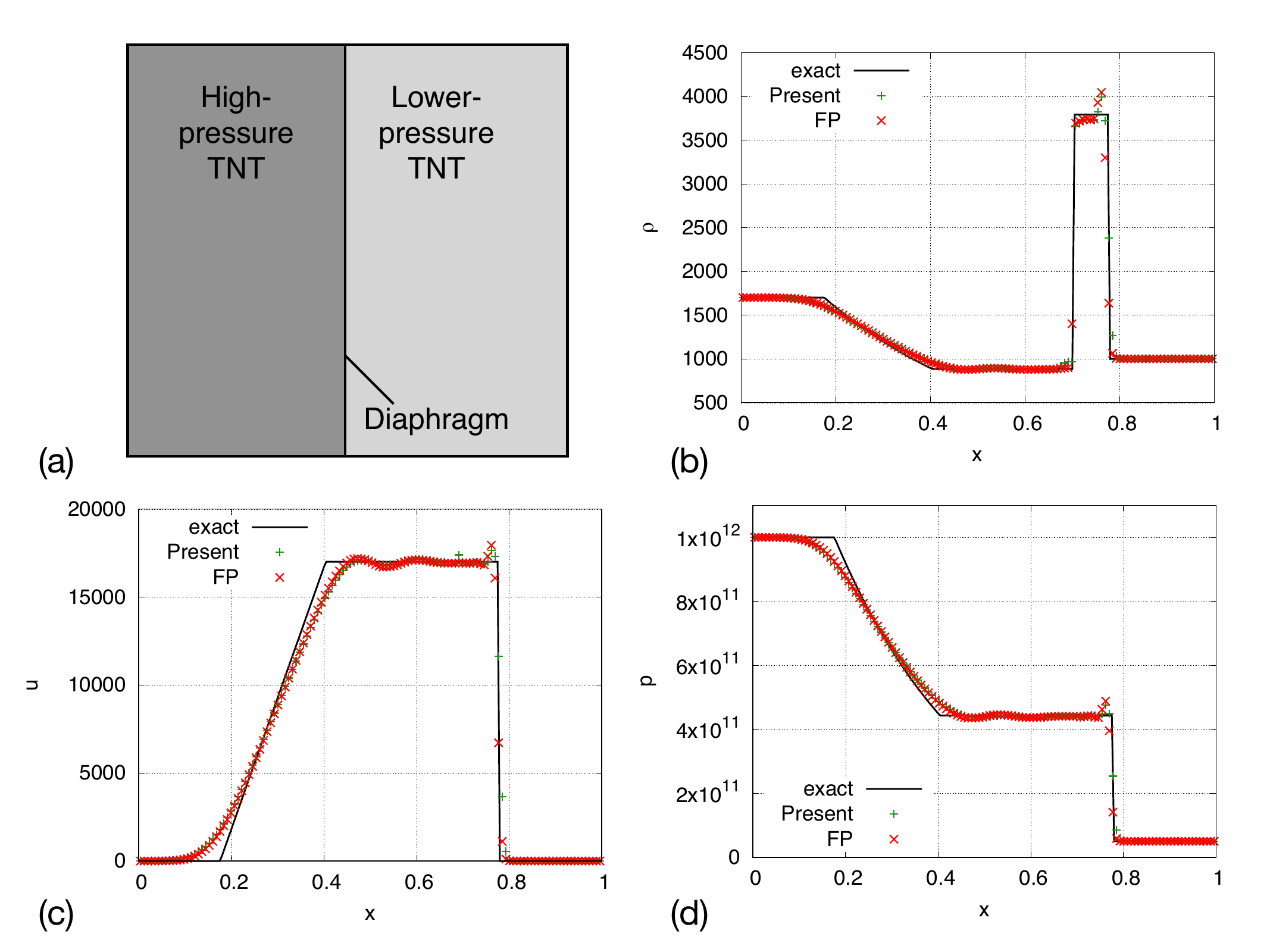}
\end{center}
\caption{Simulation setup (a) and results of (b) density, (c) velocity, and (d) pressure for the explosive TNT shock tube problem at $t=12 \mu s$. The solid lines represent the exact solution, and ``FP" represents the results using the method by Fuster and Popinet \cite{Fuster_2018a}. }
\label{fig:shocktube_jwl}
\end{figure}

To demonstrate the capability of the present methods on incorporating different equations of state, we have also considered the 1D shocktube consists of TNT explosive products \cite{Shyue_2001a,Kamm_2015a}. The JWL equation of state is employed and the JWL parameters for TNT are given in Table I of Shyue \cite{Shyue_2001a}.
 The initial conditions for the left and right states are
\begin{align}
 \{\rho, u, p\} &=
 \begin{cases}
\{1700, 0, 10^{12}\} & 0 \le x \le 0.5, \\
  \{1000, 0, 5 \times 10^{10}\} & 0.5 < x \le 1.
 \end{cases}
\end{align}
The results are compared to the FP method and exact solution in  Fig.\ \ref{fig:shocktube_jwl} at $t=12$ \textmu s. Both numerical methods compare well to the exact solution, with a reduction in the overshoot at the shock-front for the present model compared to the FP method.

\subsection{Two-phase shocktube}
\label{sec:2p_shocktube}
The 1D gas-liquid two-phase shocktube problem is employed to test the present method on capturing interfaces separating two different phases. The only change here, compared to the single-phase shocktube test (Fig.\ \ref{fig:shocktube_Sod}), is to replace the low-pressure gas by the low-pressure liquid. The problem has been used as a model to study underwater explosions \cite{Shyue_1998a,Johnsen_2006a}. The domain is a square with $L=10$ and $x=[-5,5]$. The diaphragm is initially located at $x=0$. The initial fluid properties are given as
\begin{align}
 \{\rho, u, p, \gamma, \Pi_{\infty}, f\} &=
 \begin{cases}
\{1.241, 0, 2.753, 1.4, 0, 	0\} & -5 \le x \le 0, \\
  \{0.991, 0, 3.059 \times 10^{-4}, 5.5, 1.505, 1\} & 0 \le x \le 5.
 \end{cases}
\end{align}
Consistent with former studies \cite{Shyue_1998a,Johnsen_2006a}, the viscosity and surface tension are neglected in this test. As a result, the pressure is continuous across the interface while the temperature changes abruptly. 

\begin{figure}[tbp]
\begin{center}
\includegraphics [width=1\columnwidth]{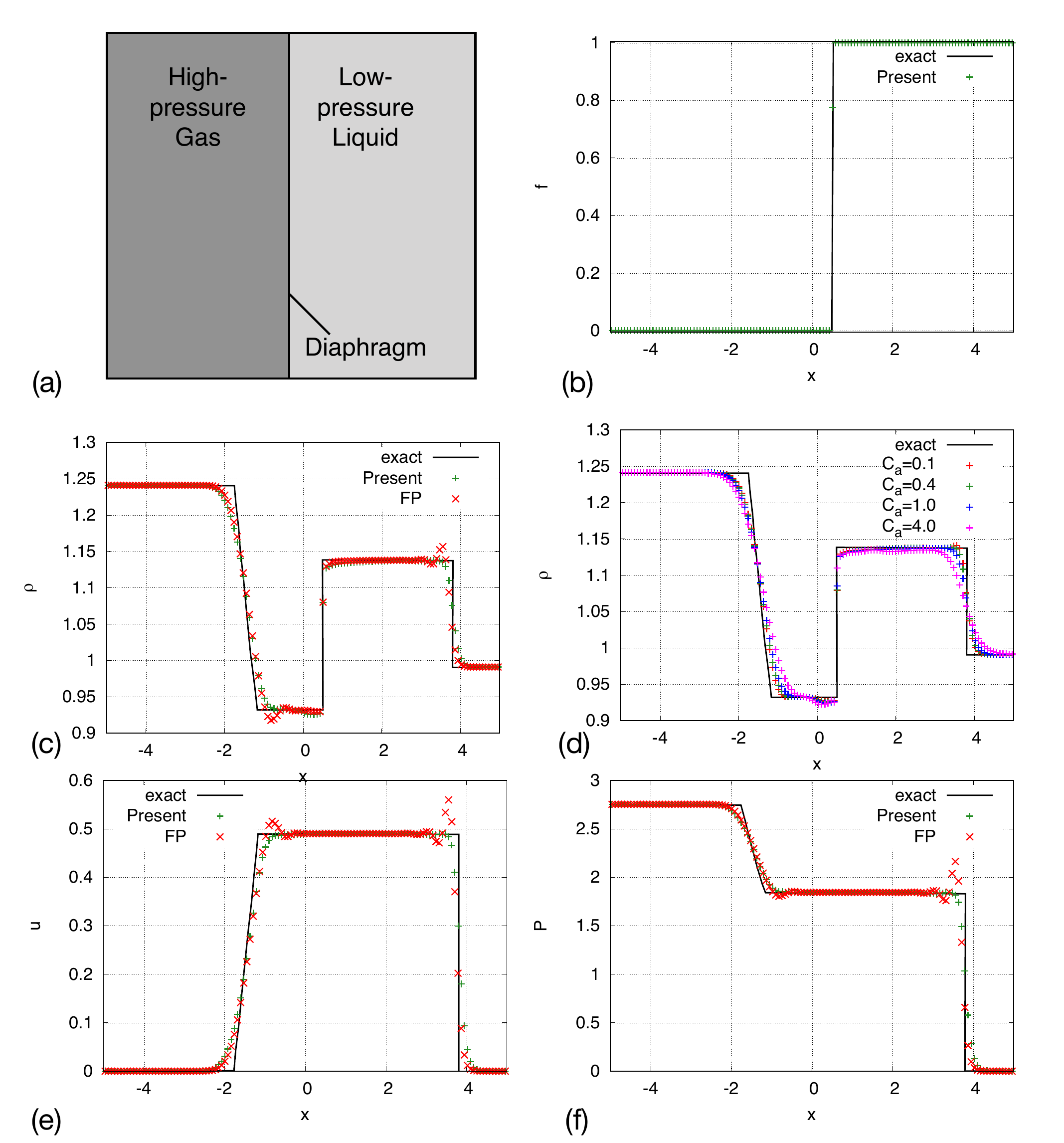}
\end{center}
\caption{Simulation setup (a) and results of (b) liquid volume fraction, (c, d) density, (e) velocity, and (f) pressure for the gas-liquid two-phase shocktube problem at $t=1$. The solid lines represent the exact solution, and ``FP" represents the results using the method by Fuster and Popinet \cite{Fuster_2018a}. Except for (d), where the acoustic Courant number ($C_a$) is varied, the rest results are for $C_a=0.4$.  }
\label{fig:shocktube_2phase}
\end{figure}

The present results for the liquid volume fraction, density, velocity, and pressure are compared with the exact solutions in Fig.\ \ref{fig:shocktube_2phase} and a good agreement is achieved. As shown in Fig.\ \ref{fig:shocktube_2phase}(b), the gas-liquid interface is captured by the VOF method as a  genuine discontinuity (with the thickness of one cell). Furthermore, the velocity and pressure are continuous at the interface without any numerical oscillations, which is an important feature that is not trivial to achieve numerically \cite{Abgrall_2001a}. 

The numerical oscillations induced by the FP method are even more profound for the two-phase shocktube test, compared to Fig.\ \ref{fig:shocktube_Sod}. Spurious oscillations are observed not only near the cylindrical shock but also near the tailing edge of the expansion fan.

The time step for the results in Fig.\ \ref{fig:shocktube_Sod} are generally calculated based on $C_a=0.4$, except for Fig.\ \ref{fig:shocktube_Sod}(d). The present all-Mach method is stable even when time steps are larger than the acoustic time step, \ie, $C_a>1$. The results for $C_a$ varying from 0.1 to 4 are shown in Fig.\ \ref{fig:shocktube_Sod}(d). The results for $C_a=0.1$ and 0.4 are almost the same. When $C_a=1$ and 4, it is observed that additional smearing is introduced at the shock and expansion fan, but the shock speeds are still correctly captured.

\subsection{Shock-bubble interaction}

\label{sec:shock-bubble}
\begin{figure}[tbp]
\begin{center}
\includegraphics [width=.9\columnwidth]{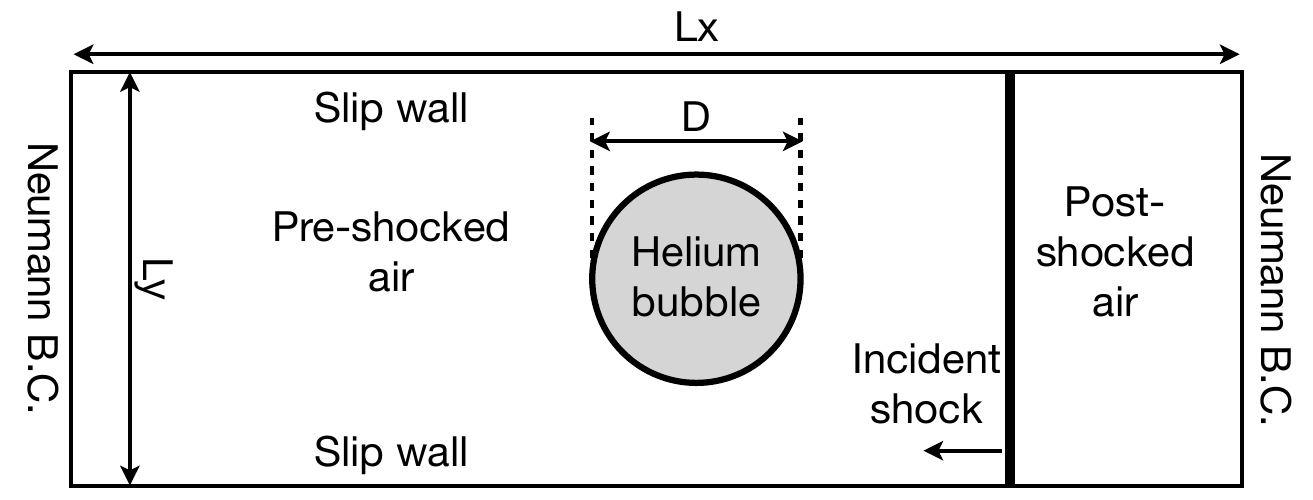}
\end{center}
\caption{Simulation setup for the shock interaction with a 2D bubble.}
\label{fig:bubble_domain}
\end{figure}

The interaction between a planar shock wave and a cylindrical helium bubble \citep{Haas_1987a} is employed to test the present numerical methods in capturing the interaction between shock and curved interfaces. The simulation results are compared with the experimental \citep{Haas_1987a} and numerical results \citep{Quirk_1996a, Terashima_2009a, Aslani_2018a} in previous studies. The computational domain and setup are shown in Fig.\ \ref{fig:bubble_domain}. The helium bubble is surrounded by air. The incident shock is coming toward the bubble from the right, with the shock Mach number $M_s=1.22$. The diameter of the bubble is $D=$50 mm. The domain width and height are $L_x=267$mm and $L_y=89$ mm, respectively. 
Both air and helium are considered as ideal gases. Given the time of interest, the physical diffusion between the two gases can be neglected. As a result, the air ($f=0$) and helium ($f=1$) are taken to be immiscible and are separated by sharp interfaces. 

The initial fluid properties are given as
\begin{align}
 \{\rho, u, p, \gamma,  f\} &=
 \begin{cases}
\{1.176, 0, 1.103\times 10^5, 1.4, 	0\} & \mathrm{Preshocked\ air}, \\
  \{1.618, -115.8, 1.591\times 10^{5}, 1.4, 0\} & \mathrm{Postshocked\ air}, \\
  \{0.219, 0, 1.013 \times 10^{5}, 1.648, 1\} & \mathrm{Helium\ bubble}, 
 \end{cases}
\end{align}
 in SI units. For the bubble size in this test, the viscous effect in the early stage of interaction is negligible, so it is neglected in the simulation. The left and right boundaries are prescribed as Neumann boundary conditions for all conservative variables, while the top and bottom boundaries are treated as slip walls. The domain is discretized by a uniform mesh. Grid refinement studies were carried out using different cell size $\Delta x$, for which $D/\Delta x=$ 71, 143 and 287. 

\begin{figure}[htbp]
\begin{center}
\includegraphics [width=1\columnwidth]{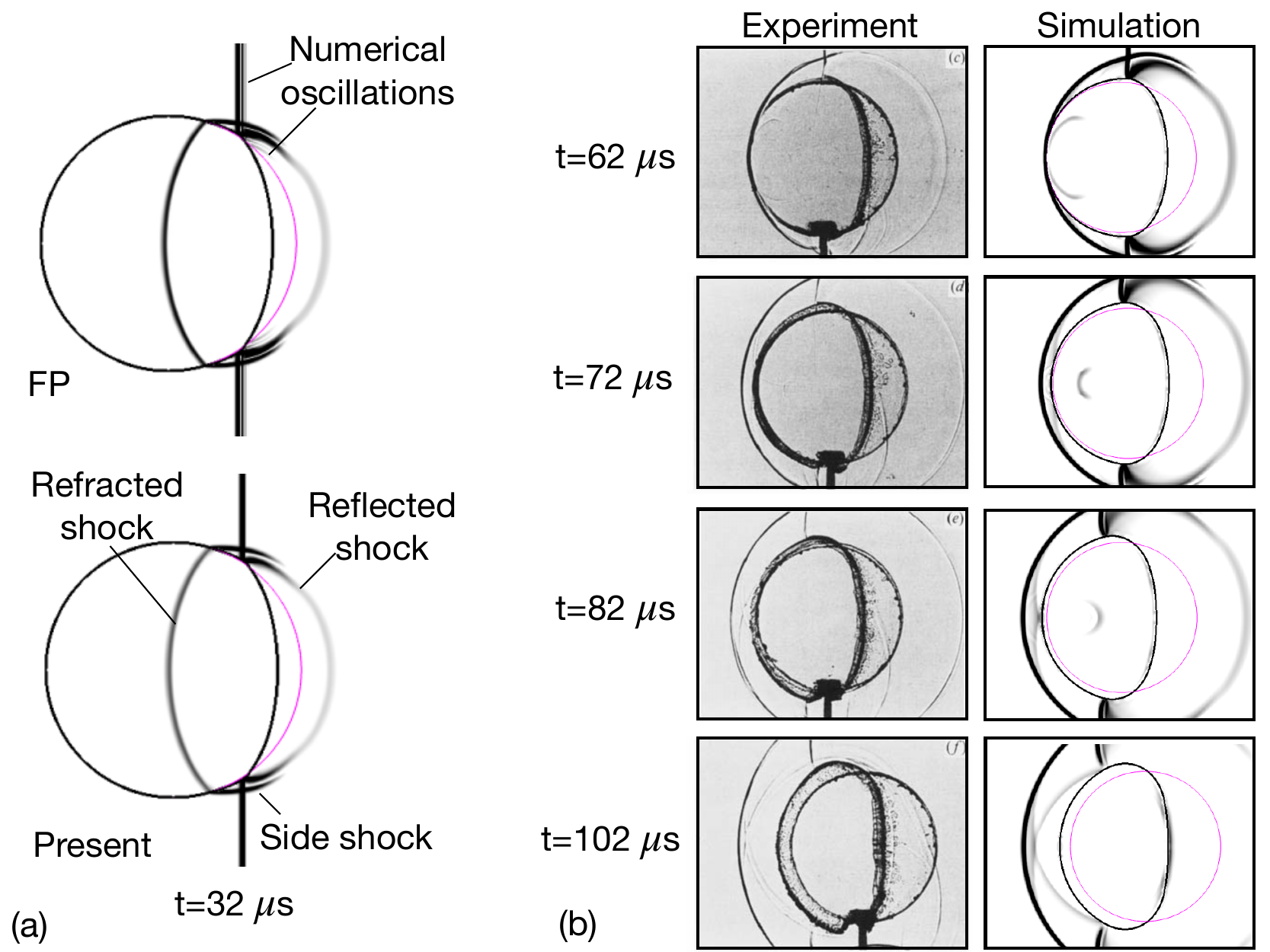}
\end{center}
\caption{(a) Comparison between the simulation results for the present and the FP methods, and (b) comparison between the present simulation results and the experimental shadowgraphs \citep{Haas_1987a}.}
\label{fig:bubble_evolution}
\end{figure}

The results obtained by the present and the FP methods for $D/\Delta x=287$ are compared in Fig. \ref{fig:bubble_evolution}(a).  It is again confirmed that the numerical oscillations observed in the FP results are successfully eliminated by the present method without smearing the interface. The waves generated in shock-bubble interaction, including the refracted, reflected, and side shocks, are well resolved by the present method. The numerical Schlieren  images (contours of density gradient) are also compared with the experimental shadowgraphs at different times in Fig. \ref{fig:bubble_evolution} (b). The temporal evolutions of the waves and the bubble interfaces predicted by the present simulation are in excellent agreement with the experimental results. 

\begin{figure}[tbp]
\begin{center}
\includegraphics [width=.9\columnwidth]{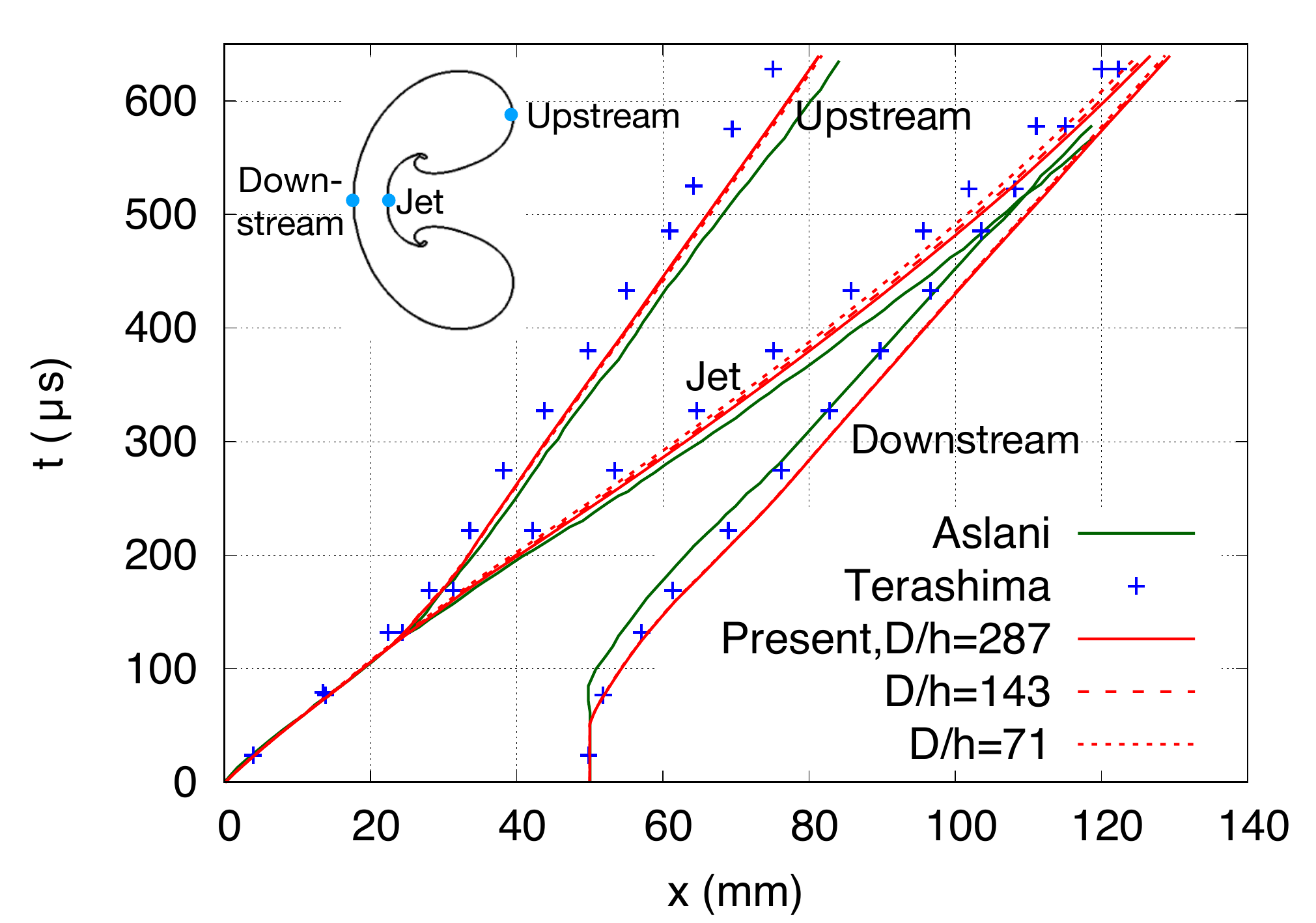}
\end{center}
\caption{Temporal evolutions of the characteristic length scales of the bubble. The present results with different mesh resolutions, indicated by different line patterns, are compared with the numerical results of Terashima and Tryggvason \cite{Terashima_2009a} and Aslani and Regele \cite{Aslani_2018a}. }
\label{fig:bubble_length}
\end{figure}

The characteristic length scales on the bubble shape, including the air-jet penetration length, and the upstream and downstream locations, are measured for quantitative validation of the present simulation results. The temporal evolutions of these characteristic length scales are shown in Fig.\ \ref{fig:bubble_length}. The present simulation results  using the fine mesh $D/\Delta x=287$ are converged. Furthermore, the present results agree  well with those by Terashima and Tryggvason \citep{Terashima_2009a} and Aslani and Regele \citep{Aslani_2018a}. The upstream location and the air-jet penetration length predicted by the present simulation results lie between the results of Terashima and Tryggvason \citep{Terashima_2009a} and Aslani and Regele \citep{Aslani_2018a}. The predicted downstream location slightly shifts to the right. The difference between the present results and others is in general small and is within the discrepancy ranges between previous numerical results \cite{Terashima_2009a}.

\subsection{Shock-droplet interaction}
\label{sec:shock-drop}
\begin{figure}[htbp]
\begin{center}
\includegraphics [width=.5\columnwidth]{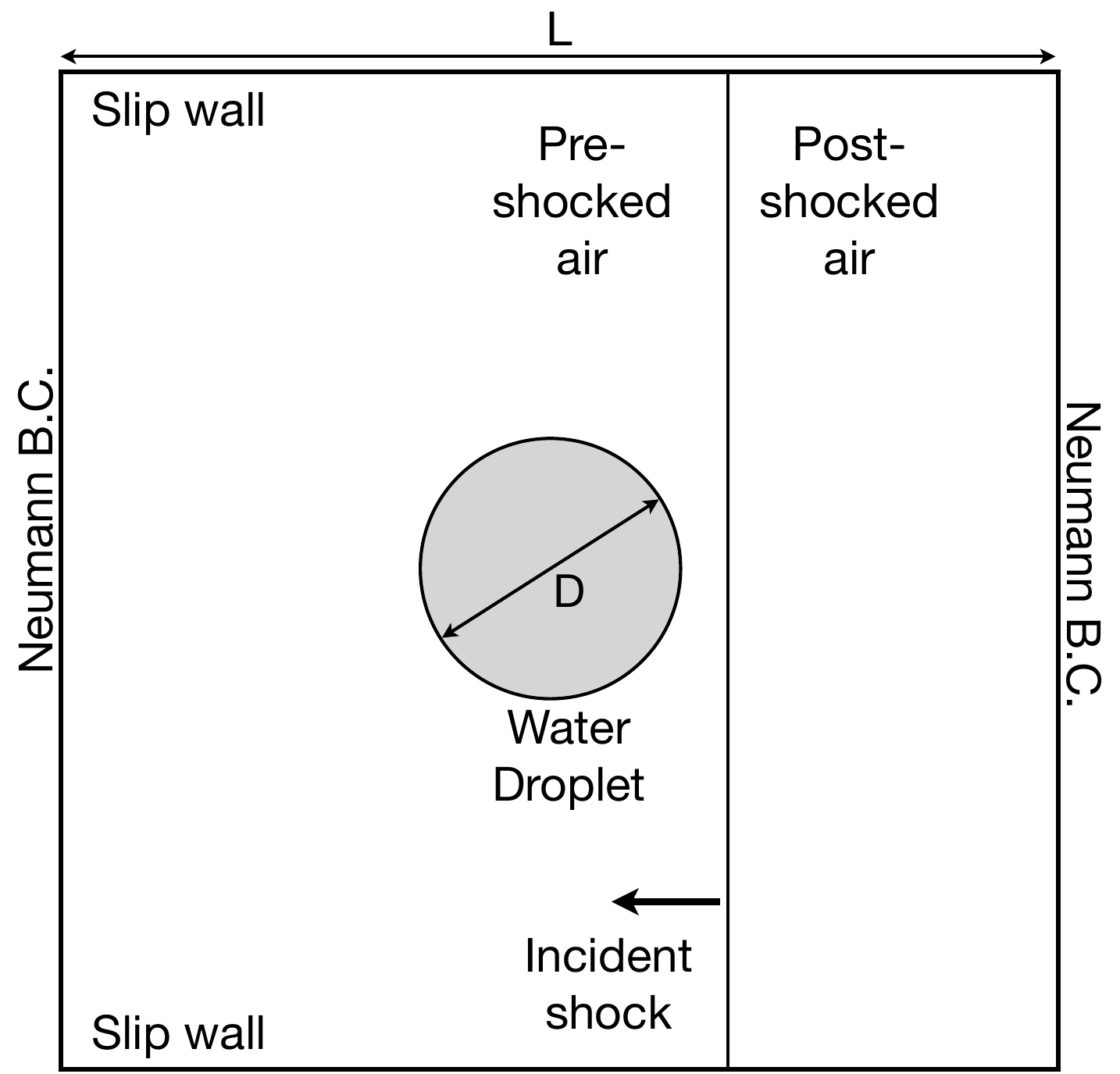}
\end{center}
\caption{Simulation setup for the shock-drop interaction.}
\label{fig:2Ddrop_domain}
\end{figure}

The interaction between a planar air shock and a liquid droplet is simulated to further examine the the present method in resolving shock interaction with curved interfaces separating two phases with significantly different fluid properties. The computational domain is a square for 2D and a cube for 3D. The droplet is initially located at the center of the domain, and a planar shock comes from right and interacts with the drop. The Neumann boundary conditions are applied to the left and right boundaries, while all lateral boundaries are treated as slip walls, similar to the shock-bubble test, see Fig.~\ref{fig:2Ddrop_domain}. 

We first consider the shock interaction with a 2D water drop (cylinder). The fluid properties and initial conditions are chosen based on the experiment of Igra and Takayama \citep{Igra_2001a}. The gas and liquid phases are air and water respectively. The drop diameter is $D=$5 mm and the incident shock Mach number is 1.47. The square domain length is taken to be $L=160$ mm and the cell size is $\Delta x=D/64$.  Since the time of interest is significantly smaller than the capillary and viscous time scales, the viscous and capillary effects are neglected. The initial conditions and fluid properties are given as follows
\begin{align}
 \{\rho, u, p, \gamma, \Pi_\infty, f\} &=
 \begin{cases}
\{1.203, 0, 1.012\times 10^5, 1.4, 	0, 0\} & \mathrm{Preshocked\ air}, \\
  \{2.176, -225.8, 2.383\times 10^{5}, 1.4, 0,0\} & \mathrm{Postshocked\ air}, \\
  \{988, 0, 1.012 \times 10^{5}, 5.5, 4.921\times 10^8,1\} & \mathrm{Water\ drop}, 
 \end{cases}
 \label{eq:shock-drop}
\end{align}
 in SI units. The values of $\gamma$ and $\Pi_\infty$ are chosen following the work of Cocchi \etal \cite{Cocchi_1996a}. 

\begin{figure}[htb]
\begin{center}
\includegraphics [width=1\columnwidth]{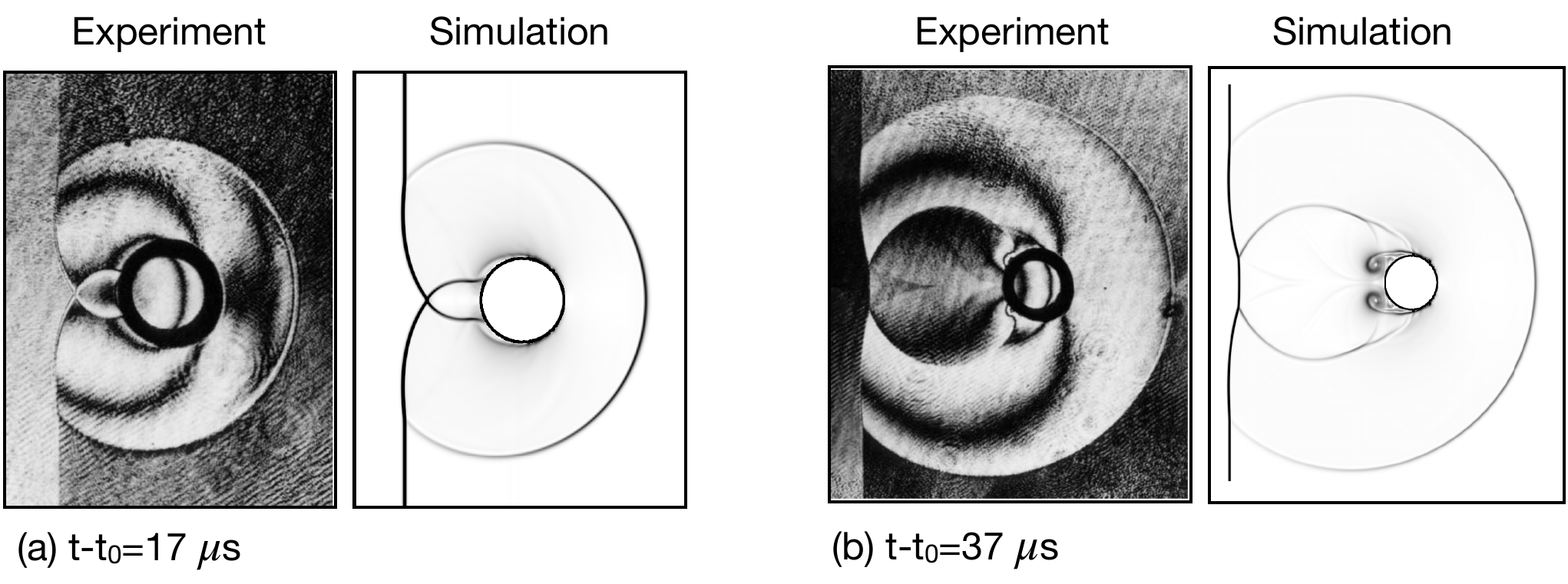}
\end{center}
\caption{Present simulation results for shock interaction with a 2D water drop, compared with the holographic interferograms of Igra and Takayama \citep{Igra_2001a}. }
\label{fig:drop_comparison}
\end{figure}

The simulation results at two different time instances are compared with the experimental holographic interferograms in Fig.~\ref{fig:drop_comparison}. The two experimental measurements were claimed to be taken at 23 and 43 \textmu s after the incident shock reaches the droplet \cite{Igra_2001a}. However, as addressed by Meng and Colonius \cite{Meng_2015a}, the times in the original experiment seem to be not calibrated properly. By examining the shock locations far away from the droplet at the two different times given in the experimental and numerical results by Igra and Takayama \citep{Igra_2001a}, we found that the shock reaches the droplet at about $t\approx6$ \textmu s, according to the zero time defined in the original paper. Therefore, we have used the simulation results at $t-t_0=17$ and $37$ \textmu s, where $t_0$ denotes the time when the shock just reaches the drop, to compare with the experimental results. It can be observed that the wave patterns arising from the shock-drop interaction predicted by the present simulation at these two time instances agree very well with the experiment. 

\begin{figure}[htb]
\begin{center}
\includegraphics [width=1.\columnwidth]{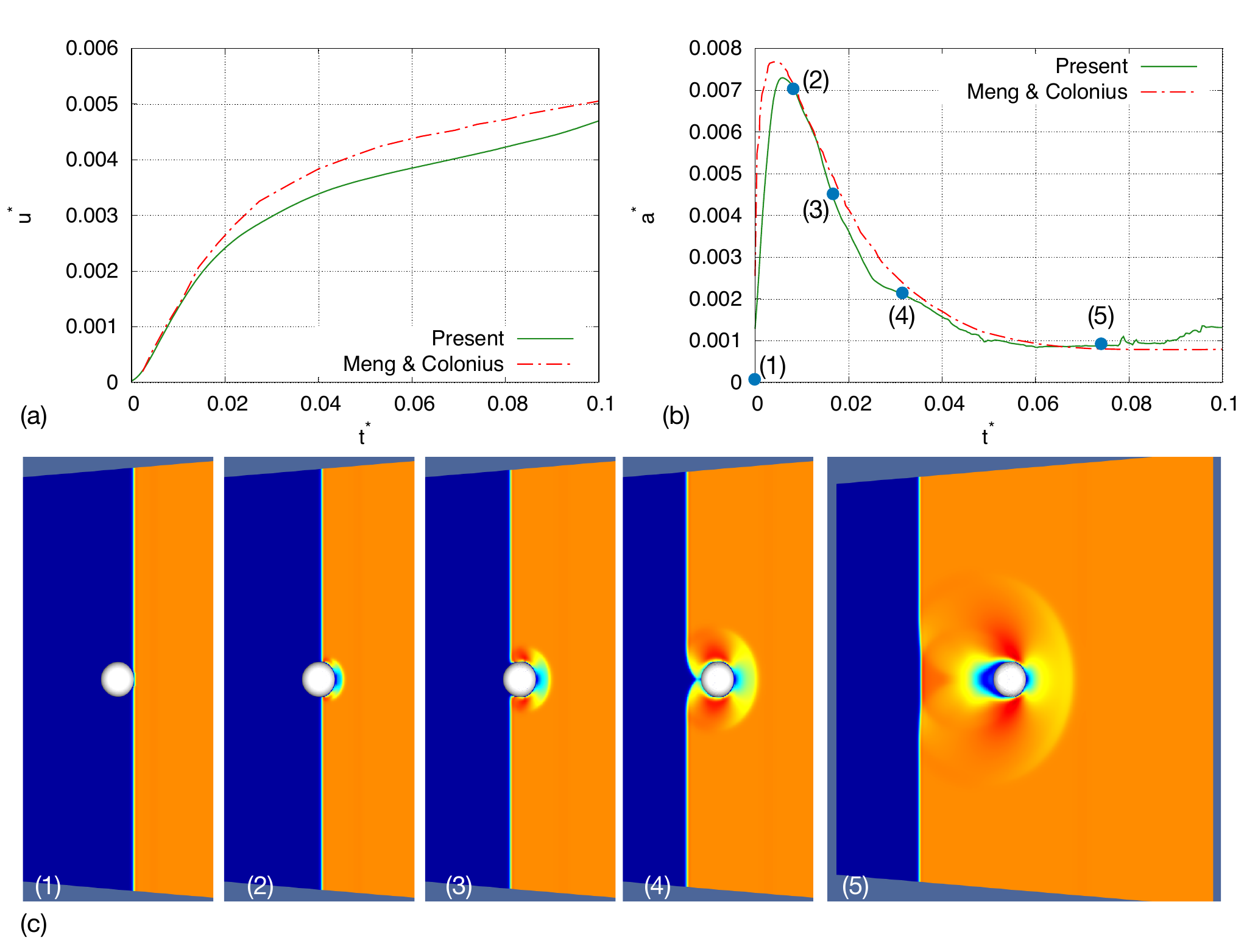}
\end{center}
\caption{3D simulation results for the shock-droplet interaction: temporal evolutions of (a) the average droplet velocity (b) acceleration, compared with the numerical results of Meng and Colunius \citep{Meng_2018a}, and (c) the velocity field on the central plane.  }
\label{fig:3d_drop}
\end{figure}

A fully 3D simulation has also been performed for the shock-droplet interaction. Similar to the 2D case, surface tension and viscosity are neglected. The drop diameter, incident shock Mach number, initial conditions, and fluid properties are the same as the 2D case. The computational domain is a cube with edge length $L=12D$. A uniform mesh (384$^3$) is used and the cell size is $\Delta x=D/32$. Due to the high computational cost, a small domain is used in this test and only a short term after the shock-drop interaction is simulated.

Following the work of Meng and Colonius \citep{Meng_2018a}, the time, the drop centroid velocity and acceleration are non-dimensionalized as $t^*=t\frac{u_{ps}}{D}\sqrt{\frac{\rho_{ps}}{\rho_l}}$, $u^{*}=u/u_{ps}$ and $a^{*}=aD/u_{ps}^2$, where $\rho_{ps}$ and $u_{ps}$ are the postshock gas density and velocity, respectively, see Eq.\ \eqr{shock-drop}. The temporal evolutions of the drop velocity and acceleration predicted by the present simulation are compared with the simulation results by Meng and Colonius \citep{Meng_2018a} in Fig.\ \ref{fig:3d_drop} (a) and (b), and a good agreement is achieved. In particular, the large acceleration induced by the passage of the shock over the droplet is well captured \cite{Ling_2011a, Ling_2013a}, see $0<t^*<0.03$ in Fig.\ \ref{fig:3d_drop}(b). The small discrepancy is probably due to the relatively low mesh resolution used in the present test. The velocity fields on the central plane at different time instants, as indicated in Fig.\ \ref{fig:3d_drop}(b), are shown in Fig.\ \ref{fig:3d_drop}(c). It can be seen that the shock refraction and reflection due to the shock-drop interaction are well captured. 

\subsection{Linear single-mode Richtmyer-Meshkov instability}
\label{sec:rmi}
As the surface tension and viscosity are ignored in the tests above, the Richtmyer-Meshkov instabilities (RMI) \cite{Richtmyer_1960a} is simulated to examine the present method in resolving the capillary and viscous effects on shock-interface interaction. The RMI is triggered by the shock interaction with a perturbed interface, which plays an essential role in the interaction between shocks and bubbles/droplets. Here we only consider the linear regime of single-mode RMI. Different Weber $We$ and Reynolds $Re$ numbers are simulated. The effects of $We$ and $Re$ on the development of linear RMI have been studied theoretically by Mikaelian \cite{Mikaelian_1990a, Mikaelian_1993a} and Carles and Popinet \cite{Carles_2002a}. DNS of RMI with different $Re$ were performed by Walchli and Thornber \cite{Walchli_2017a}. The effect of surface tension on RMI has been studied through DNS recently by Corot \etal \cite{Corot_2020a}, yet a detailed comparison against theory was not provided. 

\begin{figure}[tbp]
\begin{center}
\includegraphics [width=1.\columnwidth]{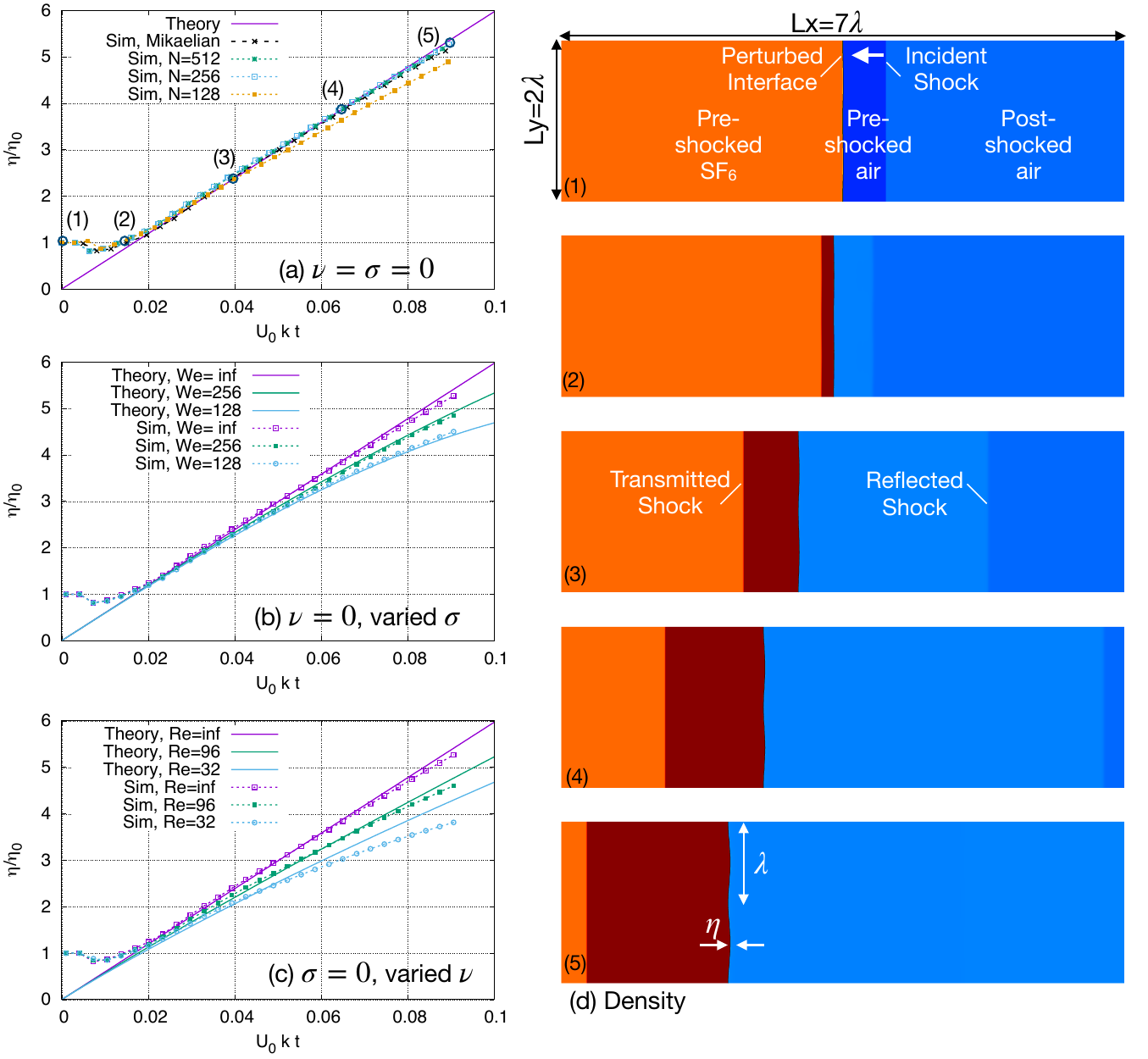}
\end{center}
\caption{Simulation results for single-mode RMI. (a) Inviscid case ($\sigma=\nu=0$) with different mesh resolutions $N=\lambda/\Delta=128$, 256 and 512, comparied with the simulation results of Mikaelian \cite{Mikaelian_1993b} and the inviscid theory of Richtmyer. (b) Results for $We=\infty, 256, 128$ and $\nu=0$ compared with the theory of Carles and Popinet \cite{Carles_2002a}. (c) Results for $Re=\infty, 96, 32$ and $\sigma=0$ compared with the theory of Carles and Popinet \cite{Carles_2002a}. (d) Temporal evolutions of the density field. The mesh resolution for (b), (c), and (d) is $N=256$.}
\label{fig:RMI}
\end{figure}

The simulation setup is shown in Fig.\ \ref{fig:RMI}. A planar air shock moves from right to left toward a perturbed interface separating air and SF$_6$. The incident shock velocity is $u_s=422.88$ m/s (shock Mach number $M_s=1.24$). Both gases are considered as ideal gases, so $\Pi_\infty=0$. The initial conditions and fluid properties are given as
\begin{align}
 \{\rho, u, p, \gamma, f\} &=
 \begin{cases}
\{1.22, 0, 1.013\times 10^5, 1.4, 	0\} & \mathrm{Preshocked\ air}, \\
  \{2.176, -123.1, 1.649\times 10^{5}, 1.4, 0\} & \mathrm{Postshocked\ air}, \\
  \{6.20, 0, 1.013 \times 10^{5}, 1.09,1\} & \mathrm{Preshocked\ SF}_6, 
 \end{cases}
\end{align}
in SI units. The wavelength and amplitude of the initial perturbation are $\lambda=3.75$ cm and $\eta_0=0.01$ cm, respectively. When surface tension and viscosities are zero, the selected parameters are the same as the simulation of Mikaelian \cite{Mikaelian_1993b}. The surface tension and viscosities are then arbitrarily varied to study the effect of $We$ and $Re$ on the development of RMI. 

The simulation results are validated against the linear theories of Mikaelian \cite{Mikaelian_1990a, Mikaelian_1993a} and Carles and Popinet \cite{Carles_2002a}. The current focus is on the early-time behavior, and the theoretical model of Carles and Popinet \cite{Carles_2002a} for the effect of surface tension yields identical results as that by Mikaelian \cite{Mikaelian_2014a}. The temporal evolution fo the perturbation amplitude is expressed as 
\begin{align}
	\frac{\eta}{\eta_0} = 1 + \Delta u k A t -  \frac{\omega^2 t^2}{2} \left(1+ \frac{\Delta u k A t}{3} \right)\,,
	\label{eq:rmi_We}
\end{align}
where $\omega=\sqrt{k^3 \sigma/(\rho_1+\rho_2)}$ is the capillary frequency.   

The theory of Carles and Popinet \cite{Carles_2002a} for the viscous effect, namely 
\begin{align}
	\frac{\eta}{\eta_0} = 1 + \Delta u k A t\left( 1 - \frac{ 16 k \sqrt{\mu_1 \mu_2 \rho_1 \rho_2}\sqrt{t}}{3\sqrt{\pi}(\sqrt{\mu_1\rho_1} + \sqrt{\mu_2 \rho_2}) (\rho_1+\rho_2)} \right)\,,
	\label{eq:rmi_Re}
\end{align}
 yields better prediction at early time, as shown in previous studies \cite{Mikaelian_2014a, Walchli_2017a}, and thus will be used here for comparison. In the inviscid limit with $\mu_1=\mu_2=\sigma=0$, Eqs.\ \eqr{rmi_We} and \eqr{rmi_Re} reduce back to the classic theory of Richtmyer \cite{Richtmyer_1960a}, namely $\eta/\eta_0=1+\Delta u k A t$. 

The post-shocked fluid densities for SF$_6$ and air are $\rho_{1b}=11.16$ kg/m$^3$ and $\rho_{2b}=1.93$ kg/m$^3$, respectively,  and the velocity change induced by the shock passage is $\Delta u=81.1$ m/s. As a result, the Atwood number $A=(\rho_{1b}-\rho_{2b})/(\rho_{1b}+\rho_{2b})(1-\Delta u/u_s)=0.570$. The Richtmyer velocity, $U_0=\eta_0 \Delta u k A=0.77$ m/s. The Weber and Reynolds numbers are defined as 
\begin{equation}
	We=\frac{(\rho_{1b}+\rho_{2b})(\Delta u)^2}{k\sigma}\,, \quad Re=\frac{\Delta u}{k\nu}\, .
\end{equation}
For convenience we simply set $\nu_1=\nu_2=\nu$ and $\lambda_v=0$. Two different $We$ and $Re$ numbers are simulated, namely $We=128$ and $256$  and $Re=32$ and $96$. The results are summarized in Fig.\ \ref{fig:RMI}. 

The present simulation results for the inviscid limit ($\nu=0,\ \sigma=0$) with different mesh resolutions (the number of cells per wavelength $N=\lambda/\Delta=128$, 256, and 512) are shown in Fig.\ \ref{fig:RMI}(a). It is observed that the simulation results converge for $N=256$ and agree well with the simulation results of Mikaelian \cite{Mikaelian_1993b} and also the theory of Richtmyer \cite{Richtmyer_1960a}. For the linear stability theory, the perturbation amplitude grows right after the impulsive acceleration is imposed, while in simulation the perturbation amplitude first decreases due to the shock compression and then grows linearly. Here the theoretical results are plotted as $\eta/\eta_0-1$ and the simulation results are shifted in time so that the two have the same starting time for the linear growth. It is measured that the computed perturbation linear growth rate $d\eta/dt = 0.74$~m/s, which is very close to the theoretical prediction, \ie., the Richtmyer velocity $U_0=0.77$~m/s. The temporal evolutions of the density field for the inviscid case and $N=256$ are shown Fig.\ \ref{fig:RMI}(d), where the transmitted and reflected shocks and the growth of the interface perturbation are observed to be well resolved. 

The simulation results for finite $We$ and $Re$ are presented in Figs.\ \ref{fig:RMI}(b) and (c). The effect of surface tension and the effect of viscosity on RMI are similar: both will contribute to reducing the growth rate over time. The smaller the $We$ or $Re$, the larger the decrease in the  growth rate. The present simulation results agree well with the theory of Carles and Popinet \cite{Carles_2002a} for the different values of $We$ and $Re$ considered here. As the theory  is valid only for small $t$, the simulation results deviate from the theoretical predictions at later time.

\subsection{Capillary oscillations of a 2D droplet}
\label{sec:droplet_oscillations}
The tests presented above are all for high-Mach-number flows. To demonstrate that the present method is also able to resolve flows with low Mach numbers, we have simulated the capillary oscillations of a 2D droplet, following the previous works \cite{Garrick_2017b,Perigaud_2005a}. Here the simulation setup is exactly the same as Perigaud and Saurel \cite{Perigaud_2005a}. A larger surface tension $\sigma=350$ N/m is used and the viscosities in both liquid and gas are taken to be zero. The EOS parameters for the liquid are $\gamma_l=2.4$ and $\Pi_{\infty,l}=10^7$ Pa,  and those for the gas are $\gamma_g=1.4$ and $\Pi_{\infty,g}=0$ Pa. The domain size is a square with edge length equal to 1 m, and is discretized by a uniform mesh with resolution $\Delta x= \Delta y = 1/128$ m. The droplet is initially placed at the center of the domain, exhibiting an elliptical shape, 
\begin{equation}
    \frac{x^2}{a^2}+\frac{y^2}{b^2}=1
\end{equation}
where $a=0.2$ m and $b=0.12$ m. The initial densities for the liquid and gas are $\rho_l=100$ kg/m$^3$ and $\rho_g=1$ kg/m$^3$, respectively.

\begin{figure}[tbp]
\begin{center}
\includegraphics [width=0.9\columnwidth]{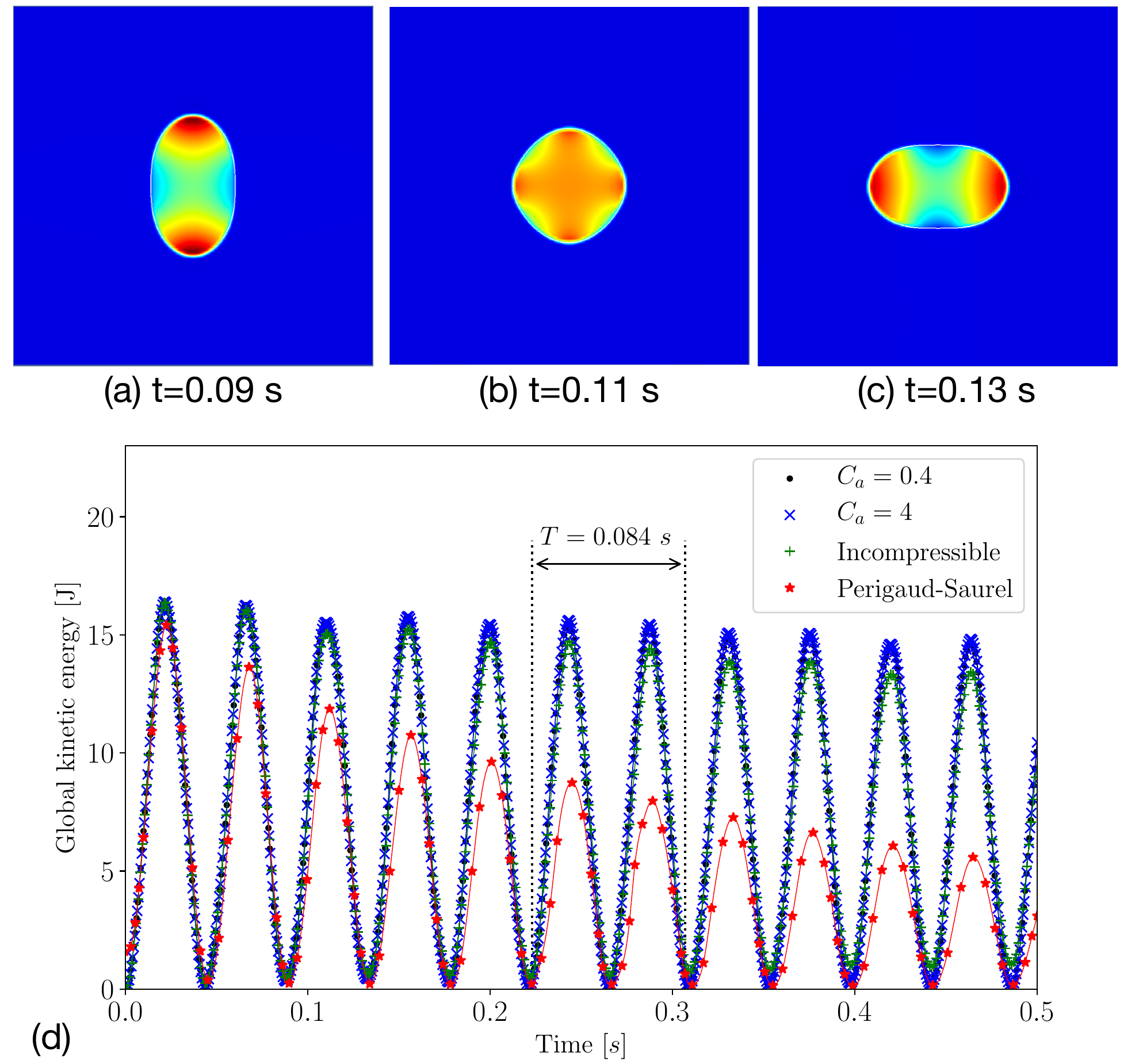}
\end{center}
\caption{Simulation results for the capillary oscillation of a 2D drop. The pressure fields and the drop surfaces at different times are shown in (a)-(c) for $C_a=0.4$. The present results for the temporal variations of the global kinetic energy for $C_a=0.4$ and 4 are shown in (d), compared with the previous numerical results by Perigaud and Saurel \cite{Perigaud_2005a} and the simulation results using the incompressible flow solver in Basilisk \cite{Basilisk, Sakakeeny_2020a}.}
\label{fig:drop_oscill}
\end{figure}

The drop surfaces and the pressure fields at $t=0.09$, 0.11, and 0.13 s are shown in Fig.\ \ref{fig:drop_oscill}, which represent different states in one oscillation cycle. The pressure variation inside the drop is due to the Laplace pressure. The dominant mode for the activated shape oscillation is clearly the second mode, for which the period is 
\begin{equation}
    T=2\pi \sqrt{\frac{(\rho_l+\rho_g)R_0^3}{6 \sigma}}
\end{equation}
where $R_0=\sqrt{ab}$ is the droplet radius at the equilibrium state. For this test case, $T=0.084$ s. 
The results for the global kinetic energy for the present method are shown in Fig.\ \ref{fig:drop_oscill}(d). Two different $\Delta t=6.4\times 10^{-6}$ and $6.4\times 10^{-5}$ s, corresponding to $C_a=0.4$ and $4$, have been used. As there are no shock waves in this problem, the KT numerical diffusion is not needed, so $D_n=0$ has been used to deactivate the numerical diffusion. The oscillation period predicted by the present simulation agrees  well with the theoretical prediction, as shown in Fig.\ \ref{fig:drop_oscill}(d).

The numerical results of Perigaud and Saruel  \cite{Perigaud_2005a} are also plotted for comparison. 
Due to numerical dissipation, the oscillation amplitude for Perigaud and Saruel's results decays rapidly over time. In contrast, the oscillation amplitude only decreases slightly, indicating the low numerical dissipation for the present method. When $C_a=4$ is used, $\Delta t$ is insufficient to resolve the acoustic waves induced by shape oscillation. Nevertheless, since the Mach number for the present case is very low, the compressibility effect is expected to be negligible. Therefore, the results for $C_a=0.4$ and 4 are almost identical, though the latter only requires 1/10 of the computational time of the former. We have also performed the same test using the incompressible flow solver in \emph{Basilisk}. The details about numerical methods and validation of the incompressible flow solver can be found in previous studies, such as \cite{Sakakeeny_2020a, Sakakeeny_2021a}. The results using the incompressible flows are found to agree very well with the present results using the all-Mach method, which affirms the asymptotic preserving feature of the present method and its capability to resolve flows of all speeds.

\section{Conclusions}
\label{sec:conclusions}
A new numerical method has been developed to simulate compressible interfacial multiphase flows (CIMF) that involve shock interaction with sharp interfaces. The geometrical volume-of-fluid (VOF) method is used to capture the interface and the conservative variables for both phases are advected in a consistent and conservative manner as the VOF advection. Numerical diffusion is introduced based on the Kurganov-Tadmor method in the region away from the interface to suppress spurious oscillations near shocks.  The contribution of pressure is incorporated using the projection method and is obtained by solving the Poisson-Helmholtz pressure equation. The balanced-force discretization method is used for the surface tension, while the height-function method is used to calculate the interface curvature. The present numerical method is tested by a sequence of CIMF problems. The simulation results for the single-phase and two-phase shocktube problems agree well with the exact solution, with spurious oscillations effectively suppressed. The interaction between a planar shock with a 2D helium bubble, a 2D water droplet, and a 3D water droplet are simulated to verify the capability of the present method in capturing the shock interaction with curved interfaces separating fluids with different properties. The complex wave structures induced by the shock-bubble and shock-droplet interactions, including the reflected, side, and refracted shocks are well captured. The present simulation results agree very well with the experimental shadowgraphs for both shock-bubble and shock-drop interactions. Quantitative validations are made by comparing the temporal evolutions of the characteristic length scales of the bubble shape and the mean velocity and acceleration of the drop during interaction with the shock. The linear single-mode Richtmyer-Meshkov instabilities for different Weber and Reynolds numbers are simulated to examine the capability of the present method in accurately capturing the capillary and viscous effects on shock-interface interactions. The simulation results are compared with linear stability theory and a good agreement has been achieved. Finally, the capillary oscillation of a 2D drop to validate the present method is resolving low-speed surface-tension driven flows. A time step that is larger than the acoustic time step has been used. The predicted oscillation period matches well with the theory. Furthermore, the present results agree with those obtained by the incompressible flow solvers and show very low numerical dissipation in the kinetic energy.

\section*{Acknowledgements}
This research was supported by the National Science Foundation (NSF \#1853193). The authors also acknowledge the Extreme Science and Engineering Discovery Environment (XSEDE)  and the Texas Advanced Computing Center (TACC) for providing the computational resources that have contributed to the research results reported in this paper. The Baylor High Performance and Research Computing Services (HPRCS) have been used to process the simulation results. We also thank Daniel Fuster for sharing his code and helpful discussions. The newly-developed methods have been implemented in the open-source multiphase flow solver \emph{Basilisk}, which is made available by St\'ephane Popinet and other collaborators.

\section*{References}

\begin{thebibliography}{10}
\expandafter\ifx\csname url\endcsname\relax
  \def\url#1{\texttt{#1}}\fi
\expandafter\ifx\csname urlprefix\endcsname\relax\def\urlprefix{URL }\fi
\expandafter\ifx\csname href\endcsname\relax
  \def\href#1#2{#2} \def\path#1{#1}\fi

\bibitem{Leveque_2002a}
R.~J. LeVeque, Finite volume methods for hyperbolic problems, Cambridge
  University Press, 2002.

\bibitem{Godunov_1959a}
S.~K. Godunov, A finite difference method for the numerical computation of
  discontinuous solutions of the equations of fluid dynamics, Mat.~Sb. 47
  (1959) 271--306.

\bibitem{Roe_1981a}
P.~L. Roe, Approximate {R}iemann solver, parameter vectors, and difference
  schemes, J.~Comput.~Phys. 43 (1981) 357--372.

\bibitem{Harten_1981a}
A.~Harten, P.~D. Lax, A random choice finite difference scheme for hyperbolic
  conservation laws, SIAM J.~Numer.~Anal. 18 (1981) 289--315.

\bibitem{Toro_1994a}
E.~F. Toro, M.~Spruce, W.~Speares, Restoration of the contact surface in the
  hll-riemann solver, Shock Waves 4 (1994) 25--34.

\bibitem{Lax_1954a}
P.~D. Lax, Weak solutions of nonlinear hyperbolic equations and their numerical
  computation, Comm.~Pure Appl.~Math.Math. 7 (1954) 159--193.

\bibitem{Rusanov_1961a}
V.~V. Rusanov, The calculation of the interaction of non-stationary shock waves
  with barriers, J.~Comput.~Math.~Phys.~USSR 1 (1961) 267--279.

\bibitem{Kurganov_2000a}
A.~Kurganov, E.~Tadmor, New high-resolution central schemes for nonlinear
  conservation laws and convection--diffusion equations, J.~Comput.~Phys. 160
  (2000) 241--282.

\bibitem{Osher_1988a}
S.~Osher, J.~A. Sethian, Fronts propagating with curvature-dependent speed:
  algorithms based on hamilton-jacobi formulations, J.~Comput.~Phys. 79 (1988)
  12--49.

\bibitem{Sussman_1999a}
M.~Sussman, A.~S. Almgren, J.~B. Bell, P.~Colella, L.~H. Howell, M.~L. Welcome,
  An adaptive level set approach for incompressible two-phase flows,
  J.~Comput.~Phys. 148 (1999) 81--124.

\bibitem{Osher_2001a}
S.~Osher, R.~P. Fedkiw, Level set methods: an overview and some recent results,
  J.~Comput.~Phys. 169 (2001) 463--502.

\bibitem{Saurel_2018a}
R.~Saurel, C.~Pantano, Diffuse-interface capturing methods for compressible
  two-phase flows, Annu.~Rev.~Fluid Mech. 50 (2018) 105--130.

\bibitem{Jain_2020a}
S.~S. Jain, A.~Mani, P.~Moin, A conservative diffuse-interface method for
  compressible two-phase flows, J.~Comput.~Phys. 418 (2020) 109606.

\bibitem{Hu_2001a}
H.~H. Hu, N.~A. Patankar, M.~Y. Zhu, {Direct numerical simulations of
  fluid-solid systems using the arbitrary Lagrangian-Eulerian technique},
  J.~Comput.~Phys. 169 (2001) 427--462.

\bibitem{Luo_2004a}
H.~Luo, J.~D. Baum, R.~Lohner, On the computation of multi-material flows using
  {ALE} formulation, J.~Comput.~Phys. 194 (2004) 304--328.

\bibitem{Corot_2020a}
T.~Corot, P.~Hoch, E.~Labourasse, Surface tension for compressible fluids in
  {ALE} framework, J.~Comput.~Phys. 407 (2020) 109247.

\bibitem{Unverdi_1992a}
S.~O. Unverdi, G.~Tryggvason, A front-tracking method for viscous,
  incompressible, multi-fluid flows, J.~Comput.~Phys. 100 (1992) 25--37.

\bibitem{Tryggvason_2001a}
G.~Tryggvason, B.~Bunner, A.~Esmaeeli, D.~Juric, N.~Al-Rawahi, W.~Tauber,
  J.~Han, S.~Nas, Y.~J. Jan, A front-tracking method for the computations of
  multiphase flow, J.~Comput.~Phys. 169 (2001) 708--759.

\bibitem{Bo_2011a}
W.~Bo, X.~Liu, J.~Glimm, X.~Li, A robust front tracking method: verification
  and application to simulation of the primary breakup of a liquid jet, SIAM
  J.~Sci.~Comput. 33 (2011) 1505--1524.

\bibitem{Hirt_1981a}
C.~W. Hirt, B.~D. Nichols, Volume of fluid ({VOF}) method for the dynamics of
  free boundaries, J.~Comput.~Phys. 39 (1981) 201--225.

\bibitem{Lafaurie_1994a}
B.~Lafaurie, C.~Nardone, R.~Scardovelli, S.~Zaleski, G.~Zanetti, Modelling
  merging and fragmentation in multiphase flows with {SURFER}, J.~Comput.~Phys.
  113 (1994) 134--147.

\bibitem{Scardovelli_1999a}
R.~Scardovelli, S.~Zaleski, Direct numerical simulation of free-surface and
  interfacial flow, Annu.~Rev.~Fluid Mech. 31 (1999) 567--603.

\bibitem{Sussman_2000a}
M.~Sussman, E.~G. Puckett, A coupled level set and volume-of-fluid method for
  computing 3d and axisymmetric incompressible two-phase flows,
  J.~Comput.~Phys. 162 (2000) 301--337.

\bibitem{Popinet_2018a}
S.~Popinet, Numerical models of surface tension, Annu.~Rev.~Fluid Mech. 50
  (2018) 1--28.

\bibitem{Johnsen_2006a}
E.~Johnsen, T.~Colonius, {Implementation of WENO schemes in compressible
  multicomponent flow problems}, J.~Comput.~Phys. 219 (2006) 715--732.

\bibitem{Meng_2015a}
J.~C. Meng, T.~Colonius, Numerical simulations of the early stages of
  high-speed droplet breakup, Shock Waves 25 (2015) 399--414.

\bibitem{Meng_2018a}
J.~C. Meng, T.~Colonius, Numerical simulation of the aerobreakup of a water
  droplet, J.~Fluid Mech. 835 (2018) 1108--1135.

\bibitem{Schmidmayer_2017a}
K.~Schmidmayer, F.~Petitpas, E.~Daniel, N.~Favrie, S.~Gavrilyuk, A model and
  numerical method for compressible flows with capillary effects,
  J.~Comput.~Phys. 334 (2017) 468--496.

\bibitem{Popinet_2009a}
S.~Popinet, An accurate adaptive solver for surface-tension-driven interfacial
  flows, J.~Comput.~Phys. 228~(16) (2009) 5838--5866.

\bibitem{Brackbill_1992a}
J.~U. Brackbill, D.~B. Kothe, C.~Zemach, A continuum method for modeling
  surface tension, J.~Comput.~Phys. 100 (1992) 335--354.

\bibitem{Chauveheid_2015a}
D.~Chauveheid, A new algorithm for surface tension forces in the framework of
  the fvcf--enip method, Eur.~Phys.~J.~B 50 (2015) 175--186.

\bibitem{Fuster_2018a}
D.~Fuster, S.~Popinet, An all-mach method for the simulation of bubble dynamics
  problems in the presence of surface tension, J.~Comput.~Phys. 374 (2018)
  752--768.

\bibitem{Renardy_2002a}
Y.~Renardy, M.~Renardy, {PROST}: a parabolic reconstruction of surface tension
  for the volume-of-fluid method, J.~Comput.~Phys. 183 (2002) 400--421.

\bibitem{Francois_2006a}
M.~M. Francois, S.~J. Cummins, E.~D. Dendy, D.~B. Kothe, J.~M. Sicilian, M.~W.
  Williams, A balanced-force algorithm for continuous and sharp interfacial
  surface tension models within a volume tracking framework, J.~Comput.~Phys.
  213 (2006) 141--173.

\bibitem{Afkhami_2008a}
S.~Afkhami, M.~Bussmann, Height functions for applying contact angles to 2d vof
  simulations, Int.~J.~Numer.~Meth.~Fluids 57 (2008) 453--472.

\bibitem{Perigaud_2005a}
G.~Perigaud, R.~Saurel, A compressible flow model with capillary effects,
  J.~Comput.~Phys. 209 (2005) 139--178.

\bibitem{Jemison_2014a}
M.~Jemison, M.~Sussman, M.~Arienti, Compressible, multiphase semi-implicit
  method with moment of fluid interface representation, J.~Comput.~Phys. 279
  (2014) 182--217.

\bibitem{Rohde_2015a}
C.~Rohde, C.~Zeiler, A relaxation riemann solver for compressible two-phase
  flow with phase transition and surface tension, Appl.~Numer.~Math. 95 (2015)
  267--279.

\bibitem{meng2016numerical}
J.~C.-C. Meng, Numerical simulations of droplet aerobreakup, Ph.D. thesis,
  California Institute of Technology (2016).

\bibitem{Garrick_2017a}
D.~P. Garrick, W.~A. Hagen, J.~D. Regele, An interface capturing scheme for
  modeling atomization in compressible flows, J.~Comput.~Phys. 344 (2017)
  260--280.

\bibitem{Garrick_2017b}
D.~P. Garrick, M.~Owkes, J.~D. Regele, A finite-volume hllc-based scheme for
  compressible interfacial flows with surface tension, J.~Comput.~Phys. 339
  (2017) 46--67.

\bibitem{Fechter_2018a}
S.~Fechter, C.-D. Munz, C.~Rohde, C.~Zeiler, Approximate riemann solver for
  compressible liquid vapor flow with phase transition and surface tension,
  Comput.~Fluids 169 (2018) 169--185.

\bibitem{Arienti_2019a}
M.~Arienti, M.~Ballard, M.~Sussman, Y.~C. Mazumdar, J.~L. Wagner, P.~A. Farias,
  D.~R. Guildenbecher, Comparison of simulation and experiments for multimode
  aerodynamic breakup of a liquid metal column in a shock-induced cross-flow,
  Phys.~Fluids 31 (2019) 082110.

\bibitem{Oomar_2021a}
M.~Y. Oomar, A.~G. Malan, R.~A.~D. Horwitz, B.~W.~S. Jones, G.~S. Langdon, An
  all-{Mach} number {HLLC}-based scheme for multi-phase flow with surface
  tension, Appl.~Sci. 11 (2021) 3413.

\bibitem{Xiao_2005a}
F.~Xiao, Y.~Honma, T.~Kono, A simple algebraic interface capturing scheme using
  hyperbolic tangent function, Int.~J.~Numer.~Meth.~Fluids 48 (2005)
  1023--1040.

\bibitem{Arrufat_2020a}
T.~Arrufat, M.~Crialesi-Esposito, D.~Fuster, Y.~Ling, L.~Malan, S.~Pal,
  R.~Scardovelli, G.~Tryggvason, S.~Zaleski, A momentum-conserving, consistent,
  volume-of-fluid method for incompressible flow on staggered grids,
  Comput.~Fluids 215 (2020) 104785.

\bibitem{Zhang_2020a}
B.~Zhang, S.~Popinet, Y.~Ling, Modeling and detailed numerical simulation of
  the primary breakup of a gasoline surrogate jet under non-evaporative
  operating conditions, Int.~J.~Multiphase Flow 130 (2020) 103362.

\bibitem{Kwatra_2009a}
N.~Kwatra, J.~Su, J.~T. Gretarsson, R.~Fedkiw, A method for avoiding the
  acoustic time step restriction in compressible flow, J.~Comput.~Phys. 228
  (2009) 4146--4161.

\bibitem{Kurganov_2001a}
A.~Kurganov, S.~Noelle, G.~Petrova, Semidiscrete central-upwind schemes for
  hyperbolic conservation laws and hamilton--jacobi equations, SIAM
  J.~Sci.~Comput. 23 (2001) 707--740.

\bibitem{Kurganov_2002a}
A.~Kurganov, D.~Levy, Central-upwind schemes for the saint-venant system, ESAIM
  Math.~Model.~Numer.~Anal. 36 (2002) 397--425.

\bibitem{Marsh_1980a}
S.~P. Marsh, LASL shock Hugoniot data, Vol.~5, University of California Press,
  1980.

\bibitem{Weymouth_2010a}
G.~D. Weymouth, D.~K.-P. Yue, Conservative volume-of-fluid method for
  free-surface simulations on cartesian-grids, J.~Comput.~Phys. 229~(8) (2010)
  2853--2865.

\bibitem{Aulisa_2007a}
E.~Aulisa, S.~Manservisi, R.~Scardovelli, S.~Zaleski, Interface reconstruction
  with least-squares fit and split advection in three-dimensional cartesian
  geometry, J.~Comput.~Phys. 225 (2007) 2301--2319.

\bibitem{Rudman_1998a}
M.~Rudman, A volume-tracking method for incompressible multifluid flows with
  large density variations, Int.~J.~Numer.~Meth.~Fluids 28 (1998) 357--378.

\bibitem{Chenadec_2013a}
V.~Le~Chenadec, H.~Pitsch, A monotonicity preserving sharp interface flow
  solver for high density ratio two-phase flows, J.~Comput.~Phys. 249 (2013)
  185--203.

\bibitem{Vaudor_2017a}
G.~Vaudor, T.~M\'enard, W.~Aniszewski, M.~Doring, A.~Berlemont, A consistent
  mass and momentum flux computation method for two phase flows. {Application}
  to atomization process, Comput.~Fluids 152 (2017) 204--216.

\bibitem{Lopez-Herrera_2015a}
J.~L{\'o}pez-Herrera, A.~Ga{\~n}{\'a}n-Calvo, S.~Popinet, M.~Herrada,
  Electrokinetic effects in the breakup of electrified jets: A volume-of-fluid
  numerical study, Int.~J.~Multiphase Flow 71 (2015) 14--22.

\bibitem{Bell_1989a}
J.~B. Bell, P.~Colella, H.~M. Glaz, A second-order projection method for the
  incompressible {Navier-Stokes} equations, J.~Comput.~Phys. 85 (1989)
  257--283.

\bibitem{Patkar_2016a}
S.~Patkar, M.~Aanjaneya, W.~Lu, M.~Lentine, R.~Fedkiw, Towards positivity
  preservation for monolithic two-way solid--fluid coupling, J.~Comput.~Phys.
  312 (2016) 82--114.

\bibitem{Basilisk}
S.~Popinet, The basilisk code., available from http://basilisk.fr/.

\bibitem{Quirk_1996a}
J.~J. Quirk, S.~Karni, On the dynamics of a shock-bubble interaction, J.~Fluid
  Mech. 318 (1996) 129--163.

\bibitem{Igra_2001a}
D.~Igra, K.~Takayama, Numerical simulation of shock wave interaction with a
  water column, Shock Waves 11 (2001) 219--228.

\bibitem{Carles_2002a}
P.~Carles, S.~Popinet, The effect of viscosity, surface tension and
  non-linearity on richtmyer--meshkov instability, Eur.~J.~Mech.~B/Fluids 21
  (2002) 511--526.

\bibitem{Sod_1978a}
G.~A. Sod, A survey of several finite difference methods for systems of
  nonlinear hyperbolic conservation laws, J.~Comput.~Phys. 27 (1978) 1--31.

\bibitem{Shyue_2001a}
K.-M. Shyue, A fluid-mixture type algorithm for compressible multicomponent
  flow with van der waals equation of state, J.~Comput.~Phys. 171 (2001)
  678--707.

\bibitem{Kamm_2015a}
J.~R. Kamm, An exact, compressible one-dimensional riemann solver for general,
  convex equations of state, Tech. Rep. LA-UR-15-21616, Los Alamos National Lab
  (2015).

\bibitem{Shyue_1998a}
K.~Shyue, An efficient shock-capturing algorithm for compressible
  multicomponent problems, J.~Comput.~Phys. 142 (1998) 208--242.

\bibitem{Abgrall_2001a}
R.~Abgrall, S.~Karni, Computations of compressible multifluids,
  J.~Comput.~Phys. 169 (2001) 594--623.

\bibitem{Haas_1987a}
J.~F. Haas, B.~Sturtevant, Interaction of weak shock waves with cylindrical and
  spherical gas inhomogeneities, J.~Fluid Mech. 181 (1987) 41--76.

\bibitem{Terashima_2009a}
H.~Terashima, G.~Tryggvason, A front-tracking/ghost-fluid method for fluid
  interfaces in compressible flows, J.~Comput.~Phys. 228 (2009) 4012--4037.

\bibitem{Aslani_2018a}
M.~Aslani, J.~D. Regele, A localized artificial diffusivity method to simulate
  compressible multiphase flows using the stiffened gas equation of state,
  Int.~J.~Numer.~Meth.~Fluids 88 (2018) 413--433.

\bibitem{Cocchi_1996a}
J.~P. Cocchi, R.~Saurel, J.~C. Loraud, Treatment of interface problems with
  godunov-type schemes, Shock Waves 5 (1996) 347--357.

\bibitem{Ling_2011a}
Y.~Ling, A.~Haselbacher, S.~Balachandar, {Importance of unsteady contributions
  to force and heating for particles in compressible flows. Part 1: Modeling
  and analysis for shock-particle interaction}, Int.~J.~Multiphase Flow 37
  (2011) 1026--1044.

\bibitem{Ling_2013a}
Y.~Ling, A.~Haselbacher, S.~Balachandar, F.~M. Najjar, D.~S. Stewart, Shock
  interaction with a deformable particle: Direct numerical simulations and
  point-particle modeling, J.~Appl.~Phys. 113 (2013) 013504.

\bibitem{Richtmyer_1960a}
R.~D. Richtmyer, Taylor instability in a shock acceleration of compressible
  fluids, Commun.~Pur.~Appl.~Math. 13 (1960) 297--319.

\bibitem{Mikaelian_1990a}
K.~O. Mikaelian, {Rayleigh-Taylor} and {Richtmyer-Meshkov} instabilities in
  multilayer fluids with surface tension, Phys.~Rev.~A 42 (1990) 7211.

\bibitem{Mikaelian_1993a}
K.~O. Mikaelian, Effect of viscosity on rayleigh-taylor and richtmyer-meshkov
  instabilities, Phys.~Rev.~E 47 (1993) 375.

\bibitem{Walchli_2017a}
B.~Walchli, B.~Thornber, Reynolds number effects on the single-mode
  richtmyer-meshkov instability, Phys.~Rev.~E 95 (2017) 013104.

\bibitem{Mikaelian_1993b}
K.~O. Mikaelian, Growth rate of the richtmyer-meshkov instability at shocked
  interfaces, Phys.~Rev.~Lett. 71 (1993) 2903.

\bibitem{Mikaelian_2014a}
K.~O. Mikaelian, Comment on {``The effect of viscosity, surface tension and
  non-linearity on Richtmyer--Meshkov instability''[Eur. J. Mech. B Fluids 21
  (2002) 511--526]}, Eur.~J.~Mech.~B/Fluids 43 (2014) 183--184.

\bibitem{Sakakeeny_2020a}
J.~Sakakeeny, Y.~Ling, Natural oscillations of a sessile drop on flat surfaces
  with mobile contact lines, Phys.~Rev.~Fluids 5 (2020) 123604.

\bibitem{Sakakeeny_2021a}
J.~Sakakeeny, Y.~Ling, Numerical study of natural oscillations of supported
  drops with free and pinned contact lines, Phys.~Fluids 33 (2021) 062109.

\end{thebibliography}

\end{document}